%% file: Draft_v0.tex
\documentclass[a4paper,11pt]{article}
\pdfoutput=1 
\usepackage{jheppub} 
\usepackage[T1]{fontenc} 
\usepackage{xcolor}
\usepackage{physics}
\usepackage{dsfont}
\usepackage{comment}
\usepackage{hyperref}
\usepackage{import}
\usepackage{svg}
\usepackage[british]{babel}
\usepackage{makecell}
\usepackage{ulem}

\newcommand{\1}{\mathds{1}}

\newcommand{\hscom}[1]{}
\newcommand{\appref}[1]{appendix \ref{#1}}

\newcommand{\figref}[1]{figure \ref{#1}}

\newcommand{\secref}[1]{section \ref{#1}}

\usepackage{hyphenat}
\title{Modular theory and symmetry resolution in hyperfinite von Neumann algebras}

\author[a]{Giuseppe Di Giulio,}
\author[b,c]{Moritz Dorband,}
\author[b]{Johanna Erdmenger,}
\author[b]{Henri Scheppach {}}

\affiliation[a]{The Oscar Klein Centre and Department of Physics, Stockholm University, AlbaNova, 106 91 Stockholm, Sweden}
\affiliation[b]{Institute for Theoretical Physics and Astrophysics and Würzburg-Dresden Cluster of Excellence ct.qmat, Julius-Maximilians-Universität Würzburg, Am Hubland, 97074 Würzburg, Germany}
\affiliation[c]{Max-Planck-Institute of Quantum Optics, Hans-Kopfermann-Straße 1, 85748 Garching, Germany}

\emailAdd{giuseppe.di-giulio@fysik.su.se}
\emailAdd{moritz.dorband@mpq.mpg.de}
\emailAdd{henri.scheppach@uni-wuerzburg.de}

\abstract{We study modular theory in hyperfinite von Neumann algebras, i.e.~in those of type II or type III, from the viewpoint of a subregion charge sector decomposition. We address this symmetry resolution by considering infinite tensor products of finite-dimensional algebras with fixed subregion charge values. An important ingredient is the combination of these algebras using direct integrals.
This allows us to obtain the symmetry-resolved modular operator, modular flow, and modular correlation functions for hyperfinite algebras. Our approach establishes a mathematical foundation for recent results on symmetry resolution and modular theory in conformal field theory.
Our analysis applies both to charges defined on a continuous range, or  on a discrete set. The latter is of interest for condensed matter theory.
Moreover, within the AdS/CFT correspondence we expect our findings to be relevant as a new ingredient for bulk spacetime reconstruction, including information from different boundary charge sectors.}

\begin{document} 
\selectlanguage{british}

\maketitle
\flushbottom

\section{Introduction}

In the past few years, operator algebras have attracted renewed attention. 
In particular, the theory of von Neumann algebras has been fruitfully applied to quantum gravity \cite{Jefferson:2018ksk,Kang:2018xqy,Kang:2019dfi, Leutheusser:2021qhd,Leutheusser:2021frk,Witten:2021jzq,Witten:2021unn,Chandrasekaran:2022cip,Chandrasekaran:2022eqq,Banerjee:2023eew,Engelhardt:2023xer,Aguilar-Gutierrez:2023odp,Banerjee:2024fmh,Crann:2024gkv} in the context of the AdS/CFT correspondence \cite{Maldacena:1997re,Witten:1998qj,GubserKlebanovPolyakov}. At the same time, scenarios of relevance in condensed matter physics, including phase transitions \cite{Basteiro:2024cuh} and topological phases of matter \cite{Ogata:2020ofz,Kawahigashi:2021hds}, have also been studied under the lens of the theory of operator algebras.
Originally developed in the works \cite{Murray1937,vNeumann1939,vNeumann1940,Murray1943,Takesaki1979}, von Neumann operator algebras have been historically used to rigorously frame quantum field theories (QFT) \cite{Haagbook,Arakibook}.
Modular theory is certainly one of the most successful results of the interplay between operator algebras and theoretical physics \cite{Brattelibook,Haagbook}, leading to applications to QFT \cite{Blanco:2013lea,Faulkner:2016mzt,Ceyhan:2018zfg,Lashkari:2018nsl,Lashkari:2019ixo,Mintchev:2022fcp,Jovanovic:2025mwe,Mintchev:2025yso} and quantum information \cite{Araki:1976zv,Casini:2008cr,Witten:2018zxz}. Among these results, Araki's definition of relative entropy through modular theory \cite{Araki:1976zv} has been one of the first steps towards an information-theoretical understanding of QFTs. More concretely, modular theory allows for the construction of the modular operator, a rigorously defined counterpart of reduced density matrices, which are mathematically ill-defined in QFTs without regulators.

In this work, we study aspects of modular theory related to the presence of symmetries in the system. In particular, we investigate how the modular operator and its correlation functions receive contributions from different symmetry sectors.
Analogous questions have been addressed regarding entanglement measures, leading to the notion of {\it symmetry-resolved entanglement}. This fine-grained description of the entanglement structure \cite{LaFlorencie2014,Goldstein:2017bua,Xavier:2018kqb,Lukin19,Azses:2020tdz,Vitale:2021lds,Neven:2021igr} (see \cite{Castro-Alvaredo:2024azg} for a review) has highlighted new properties of many-body quantum systems and QFTs. When symmetry-resolved entanglement entropies are computed in conformal field theory (CFT), the results depend on additional aspects of the model, beyond the sole central charge, for instance, the compactification radius for a compact boson CFT or the dimensionality of the symmetry group \cite{Bonsignori:2019naz,Murciano:2020vgh,Bonsignori:2020laa,Zhao:2020qmn,Capizzi:2020jed,Estienne:2020txv,Horvath:2021fks,Capizzi:2021kys,Calabrese:2021wvi,Foligno:2022ltu,Ghasemi:2022jxg,Capizzi:2023bpr,Fossati:2023zyz,Kusuki:2023bsp}. Moreover, it was found that, under very general circumstances, the entanglement entropy takes the same value in all symmetry sectors, a fact known as {\it entanglement equipartition} \cite{Xavier:2018kqb,Murciano:2019wdl,Fraenkel:2019ykl,Calabrese:2020tci,Pal:2020wwd,Magan:2021myk,DiGiulio:2022jjd,Ares:2022hdh,Northe:2023khz,Benedetti:2024dku}.
Apart from entanglement entropies, other quantities have been analysed in the framework of symmetry resolution, including negativity, relative entropy, generalized entropies, operator entanglement and distance between density matrices \cite{Murciano:2021djk,Cornfeld:2018wbg,Chen:2021nma,Chen:2021pls,Chen:2022gyy,Capizzi:2021zga,Parez:2022sgc,Parez:2022xur,Rath:2022qif,Ares:2022gjb,Murciano:2023ofp,Yan:2024rcl}.
Among these investigations, in \cite{Di_Giulio_2023}, the modular operator and the corresponding modular flow of invariant operators were studied using the toolkit of symmetry resolution. In that work, the authors focused on a specific type of von Neumann algebras, named type I, which allows for states with a finite amount of entanglement.
Thus, the results of \cite{Di_Giulio_2023} apply to QFTs only after a lattice regularization or by using the split property yielding a type I algebra \cite{Doplicher:1984zz}.

In the present work, we extend the analyses of \cite{Di_Giulio_2023} to {\it hyperfinite algebras}, i.e.~to those of type II or III.   Hyperfinite algebras are of fundamental importance due to their connection to QFT. For instance, as described in \cite{Haagbook}, it is generically assumed that the algebra associated to local subregions of a QFT is described by a type III$_1$ von Neumann algebra. Furthermore, when including gravitational effects, it was shown that the algebra becomes type II$_1$ \cite{Witten:2021unn,Chandrasekaran:2022cip}. Thus, our findings can be fruitfully applied in these fields in the future. 
To incorporate both type II and type III algebras, we work in the setup developed in  \cite{Powers1967,Araki1968} and reviewed in \cite{Witten:2018zxz}. In this framework, all the types of algebras (up to isomorphisms) can be described in terms of (restricted) infinite tensor products of finite-dimensional algebras and accessed by properly tuning a set of parameters. By introducing a subregion charge, we realize that the only meaningful way to build an invariant algebra that may be analysed under the lens of symmetry resolution is by combining algebras defined at fixed subregion charge. We carry out this task using direct integrals, which, to our knowledge, have never been used for symmetry resolution.

Utilizing the direct integrals, we achieve a resolution of hyperfinite von Neumann algebras into subregion charge sectors. Subsequently, we use this approach to perform the symmetry resolution of the modular operator, the modular flow of invariant operators, and the modular correlation functions. The structure of the equations follows the one found for type I algebras. 
One of the most insightful consequences of our findings is that, also for hyperfinite algebras, the symmetry-resolved modular correlation functions satisfy the Kubo-Martin-Schwinger condition if and only if the total one does.
Thus, our analysis generalizes the results of \cite{Di_Giulio_2023} to hyperfinite von Neumann algebras and strengthens their applicability in QFT. In particular, thanks to our analysis, the explicit computation in \cite{Di_Giulio_2023} for a massless Dirac CFT in the presence of a UV regulator finds formal justification in a cutoff-free QFT. 
These results open the possibility of using boundary charge sectors as new tools within the bulk reconstruction program \cite{Hamilton:2006az,Kabat:2011rz,Faulkner:2017vdd,Faulkner:2018faa}, for instance, when applied in the presence of charged black holes. Moreover, they provide a solid mathematical framework for symmetry resolution of the entanglement by discussing the modular operator decomposed into charge sectors as the algebraic counterpart of the fixed-charge reduced density matrix. 

To summarize, the main results of this work are the following: First, we construct hyperfinite algebras resolved into subregion charge sectors through direct integrals, generalizing the results of \cite{Di_Giulio_2023}. Second, we construct symmetry-resolved modular operators and modular flow. Third, we show that the symmetry-resolved modular correlation functions satisfy the KMS condition if the total modular correlation function does.

The paper is organized as follows. In \secref{sec:setuptools}, we review the main tools that we need for our investigations. Next, we consider local and total charge operators in the infinite tensor product algebras in \secref{sec:firstapproach}. The outcome of this analysis leads to the direct integral construction of invariant local hyperfinite algebras in \secref{sec:hyeprfiniteresolved}.
In this framework, in \secref{sec:modflow_Resolu}, we discuss the symmetry resolution of modular operators, modular flows of invariant operators, and modular correlation functions. Conclusions and future perspectives are provided in \secref{sec:conclusions}, while technical details are reported in the two appendices.

\section{Setup and tools}
\label{sec:setuptools}

\subsection{Review of modular theory} \label{sec:modular_theory}

We consider a net of local algebras $\mathcal{F}$ that defines a quantum theory or a QFT. More precisely, we consider $\mathcal{F}$ as a representation of the algebra on a Hilbert space $\mathcal{H}$. Furthermore, consider the collection of von Neumann algebras $\mathcal{F}(V)\subset\mathcal{F}$ associated to causally complete subregions $V$, which we denote \textit{field algebras}.
If the system at hand has a $U(1)$ symmetry, we can also define the algebra of invariant operators. If the symmetry is generated by an operator $Q\in \mathcal{F}$, this invariant algebra can be characterized as the algebra of operators in $\mathcal{F}$ which commute with $Q$. 
Furthermore, the Hilbert space $\mathcal{H}$ decomposes as \cite{Fewster:2019ixc}
\begin{align}
\label{eq:SSSdecomposition}
    {\cal H}=\bigoplus_q{\cal H}_q\,,
\end{align}
where $\mathcal{H}_q$ contains eigenvectors of $Q$ with eigenvalue $q$, which we call superselection sectors (SSS). As the invariant operators commute with $Q$, this decomposition implies that there is a representation of the algebra of invariant operators on the $q$th SSS, which we denote by
\begin{equation}
    \pi_q (\mathcal{A}) = P_q  \{a\in \mathcal{F}\,\vert \, [a,Q]=0\} P_q\,.
\end{equation}
Here, $P_q$ are the projectors onto $\mathcal{H}_q$. The representations $\pi_q (\mathcal{A})$ are called GNS representations.
Similarly, an algebra of invariant operators associated with a subregion $V$ is defined by
\begin{equation}
    \pi_q (\mathcal{A}(V)) = P_q  \{a\in \mathcal{F}(V)\,\vert \, [a,Q]=0\} P_q\,.
\end{equation}
Thus, from an algebraic perspective, we can characterize the $\mathcal{H}_q$ as the subspaces of $\mathcal{H}$ that are invariant under the action of $\pi_q (\mathcal{A})$.

Typically, the vacuum is invariant under the symmetry, and thus the vacuum sector corresponds to the eigenvalue $q=0$ or equivalently the SSS $\mathcal{H}_0$. We are mostly interested in the invariant algebra in this vacuum sector. For readability, we drop the subscript for this sector and denote
\begin{equation}\label{eq:def_invariant_algebras}
    \mathcal{A} \equiv \pi_0(\mathcal{A})\,,\qquad
    \mathcal{A}(V) \equiv \pi_0(\mathcal{A} (V))\,.
\end{equation}
We thus have the collection $\mathcal{A}(V)\subset \mathcal{A}$ and we call it \textit{algebra of observables}. We stress that this choice of the SSS is purely conventional and the following discussion can be straightforwardly extended to general SSS by reintroducing the label $q$ at the appropriate places.
Although we assume here that the symmetry is generated by a $U(1)$ charge, analogous statements as presented here exist for non-Abelian symmetry groups, which can be found in \appref{app:non_abelian_and_discrete_resolution}.

In order to define modular theory on the algebra $\mathcal{A}(V)$, consider a cyclic separating state $\ket{\Psi}\in\mathcal{H}_0$  \cite{Arakibook}. We stress that this implies that the cyclic separating state is an eigenstate of the charge $Q$
\begin{equation}
\label{eq:eigenstatecharge}
    Q\ket{\Psi}=0\,.
\end{equation}
This happens, for instance, if $\ket{\Psi}$ is the vacuum state of a $U(1)$-invariant system.
The starting point of modular theory is noticing that there exists a unique antilinear {\it modular involution} $S_\Psi$ such that \cite{Brattelibook}
\begin{equation}
\label{eq:Modular relation general}
S_\Psi a |\Psi\rangle= a^\dagger|\Psi\rangle,
\qquad
\forall a\in \mathcal{A}(V).
\end{equation}
The polar decomposition
\begin{equation}
\label{eq:polardecompositiongeneral}
S_\Psi=J_\Psi \Delta^{1/2}_\Psi
\end{equation}
allows to introduce the antiunitary modular conjugation $J_\Psi$ and the positive, self-adjoint {\it modular operator} $\Delta_\Psi$. We stress that the modular operator is identified by two ingredients: a local algebra $\mathcal{A}(V)$ and a cyclic and separating state $\ket{\Psi}$. A fundamental result of modular theory is the Tomita-Takesaki theorem \cite{Haagbook}
\begin{equation}
\label{eq:TTtheorem}
\Delta_\Psi^{\mathrm{i}t}\mathcal{A}(V)\Delta_\Psi^{-\mathrm{i}t}=\mathcal{A}(V),\qquad
\forall t \in \mathbb{R}\,.
\end{equation}
For any $t\in \mathbb{R}$ and $ a\in\mathcal{A}(V)$, $\Delta_\Psi^{\mathrm{i}t}a\Delta_\Psi^{-\mathrm{i}t}$ is called {\it modular flow}. The theorem (\ref{eq:TTtheorem}) can be rephrased by saying that the modular flow preserves the algebra $\mathcal{A}(V)$. From the modular flow, we can then construct \textit{modular correlation functions}
\begin{equation}
  \label{eq:modcorrfunc_AQFT}
    G_{\textrm{\tiny mod}}(a,b;t)
    \equiv
\langle\Psi|b\Delta_\Psi^{\mathrm{i}t}a\Delta_\Psi^{-\mathrm{i}t}|\Psi\rangle,
    \qquad
    a,b\in\mathcal{A}(V),
    \quad
    t\in\mathbb{R}\,.
\end{equation}
Regarded as a function in $t$, modular correlation functions are analytic in the strip $-1<\textrm{Im}(t)<0$, and can be analytically continued to $0<\textrm{Im}(t)<1$ by defining
\begin{equation}
\label{eq:KMS_AQFT}
 G_{\textrm{\tiny mod}}(a,b;t+\mathrm{i}) 
 =
\langle\Psi|\Delta_\Psi^{\mathrm{i}t}a\Delta_\Psi^{-\mathrm{i}t}b|\Psi\rangle\,.
\end{equation}
This continuation can be understood as resulting from the {\it Kubo-Martin-Schwinger (KMS) condition} \cite{Haagbook}, which  is a crucial feature of thermal states. Thus, the analyticity of the modular correlation function can be interpreted as the fact that, 
due to the presence of entanglement, the pure state restricted to a subregion effectively behaves as a thermal state. In this spirit, we may think of the modular flow as a time evolution of an operator with respect to the Hamiltonian $-K =\ln \Delta_\Psi$ in a thermal state with temperature $-1$, with $K$ the {\it modular Hamiltonian}.

\subsection{Symmetry-resolved modular theory in type I algebras}\label{sec:type_I_resolution}

Type I algebras arise e.g.~as algebras of bounded operators on finite-dimensional Hilbert spaces. They admit features that are typically found in quantum mechanics, such as traces and reduced density matrices. Moreover, in type I finite-dimensional algebras, the entanglement entropy between subsystems remains finite and the Hilbert space  factorizes into Hilbert spaces corresponding to local subsystems. This implies that modular theory simplifies for these algebras. This also holds for type I$_\infty$ algebras, although the Hilbert space is infinite-dimensional in this case.
In particular, a relation between modular operator and reduced density matrices emerges, which highlights the importance of the modular operator from a quantum information perspective.

For type I algebras, we may consider a Hilbert space $\mathcal{H}$ that factorizes as 
\begin{equation}
    \label{eq:hilbert_spatial_decomp}\mathcal{H}=\mathcal{H}_V\otimes \mathcal{H}_{V'}\, .    
\end{equation}
The local algebra $\mathcal{A}(V)$ of operators 
located in the region $V$ acts non-trivially only on $\mathcal{H}_V$, while the algebra $\mathcal{A}(V')$ of operators in the complementary region $V'$ only on $\mathcal{H}_{V'}$. In other words, $a\in\mathcal{A}(V)$ is represented on $\mathcal{H}$ as $a\otimes\boldsymbol{1}_{V'}$, while $a'\in\mathcal{A}(V')$ as $\boldsymbol{1}_{V}\otimes a'$. 
 According to  standard algebraic QFT references \cite{Haagbook}, the modular operator reads
\begin{equation}
\label{eq:modopRDM0}
\Delta_\Psi=
\rho_{V}\otimes\rho_{V'}^{-1}
,
\end{equation}
while the modular flow is
\begin{equation}
\label{eq:modflowRDM}
\Delta_\Psi^{\mathrm{i}t}a\Delta_\Psi^{-\mathrm{i}t}
=\rho_{V}^{\mathrm{i}t}a\rho_{V}^{-\mathrm{i}t}\otimes \boldsymbol{1}_{V'}
\,,
\end{equation}
which makes manifest the Tomita-Takesaki theorem \eqref{eq:TTtheorem}.
Note that the above relations between modular operator and reduced density matrices are only valid in the type I case.

We assume now that the generator $Q$ of the $U(1)$ symmetry can be decomposed as $Q=Q_V\oplus Q_{V'}$, where $Q_V$ is the charge restricted to the spatial subregion $V$ and
$Q_{V'}$ to the complement. The property \eqref{eq:eigenstatecharge} and the fact that, in the type I setting, we can decompose the Hilbert space into contributions from different spatial regions imply that
\begin{equation}
\label{eq:chargesectors}
   \mathcal{H}_0=\bigoplus_{\bar{q}} \mathcal{H}_{\bar{q}}^{(V)}\otimes \mathcal{H}^{(V')}_{-\bar{q}} \,,
\end{equation}
where $\bar{q}$ are the eigenvalues of $Q_V$ and the eigenvalue of $Q_{V'}$ in the second factor is chosen such that the total charge takes a fixed value, in this case $0$. We emphasize that this decomposition is different from superselection as the $\bar{q}$ are eigenvalues of the subregion charge operator $Q_V$ instead of the total charge $Q$. The state $\ket{\Psi}$ allows for a decomposition of the form
\begin{equation}
\label{eq:Schmidt_dec}
|\Psi\rangle=\sum_{\bar{q}}\sqrt{p(\bar{q})}|\Psi_{\bar{q}}\rangle\,,
\qquad
|\Psi_{\bar{q}}\rangle\in \mathcal{H}^{(V)}_{\bar{q}} 
\otimes\mathcal{H}^{(V')}_{-\bar{q}} \, .
\end{equation}
Imposing that the vectors $|\Psi_{\bar{q}}\rangle$ are normalized, the function $p(\bar{q})$ may be interpreted as the probability of finding a charge $\bar{q}$ in the region $V$. 
This distribution is non-trivial given that, although the total charge in the system is fixed (to be vanishing, in this case) and conserved, the one associated with spatial subregions can fluctuate. Moreover, \eqref{eq:Schmidt_dec} tells that the probability of measuring charge $\bar{q}$ in the complement $V'$ is given by $p(-\bar{q})$.
In what follows, we call the sectors labelled by $\bar{q}$ in \eqref{eq:chargesectors} subregion charge sectors, or, shortly, charge sectors. These sectors must not be confused with the SSS discussed in \eqref{eq:SSSdecomposition}.
To access the various charge sectors, we introduce the projectors
\begin{equation}
\label{eq:projectors_chargesector}
   \Pi_V(\bar{q}):  \mathcal{H}_0\quad\longrightarrow\quad\mathcal{H}^{(V)}_{\bar{q}}\otimes \mathcal{H}^{(V')}_{-\bar{q}} \,.
\end{equation}
We may thus write
\begin{equation}
    \label{eq:prob_def}p(\bar{q})=\bra{\Psi} \Pi_V(\bar{q})\ket{\Psi}.
\end{equation}
This retrospectively justifies the interpretation of the coefficients in the decomposition \eqref{eq:Schmidt_dec} as probabilities associated to a projective measurement.

Given the algebra of observables $\mathcal{A}(V)$, the projectors (\ref{eq:projectors_chargesector}) allow for a decomposition as
\begin{equation}
\label{eq:SRdec_algebra}
    \mathcal{A}(V)=
    \bigoplus_{\bar{q}} \mathcal{A}_{\bar{q}}(V)\,,
    \qquad \mathcal{A}_{\bar{q}}(V)\equiv \Pi_V(\bar{q})\mathcal{A}(V)\Pi_V(\bar{q})\,.
\end{equation}
It has been shown in \cite{Di_Giulio_2023} that, given the modular relation \eqref{eq:Modular relation general} associated to an algebra of observables $ \mathcal{A}(V)$ 
and a cyclic and separating state satisfying \eqref{eq:eigenstatecharge}, for any charge sector labelled by $\bar{q}$, it holds that
\begin{equation}
    \label{eq:resolved_Modular relation}
S_{\Psi,\bar{q}} a_{\bar{q}} |\Psi_{\bar{q}}\rangle= a_{\bar{q}}^\dagger|\Psi_{\bar{q}}\rangle,
\qquad
\forall  a_{\bar{q}}\in \mathcal{A}_{\bar{q}}(V)
    \,,
\end{equation}
where
\begin{equation}\label{eq:SR_Tomita_and_csstate}
    S_{\Psi,\bar{q}}\equiv S_\Psi\Pi_V(\bar{q})\,,
    \qquad\qquad
    |\Psi_{\bar{q}}\rangle\equiv\frac{\Pi_V(\bar{q})|\Psi\rangle}{\sqrt{p(\bar{q})}}\,.
\end{equation}
This equation can be regarded as a modular relation associated to the local algebra $\mathcal{A}_{\bar{q}}(V)$ and to the state $|\Psi_{\bar{q}}\rangle\in\mathcal{H}^{(V)}_{\bar{q}}\otimes \mathcal{H}^{(V')}_{-\bar{q}} $. Indeed, it was shown that, if $|\Psi\rangle$ is cyclic and separating for $\mathcal{A}(V)$ of type I, then $|\Psi_{\bar{q}}\rangle$ is cyclic and separating for $\mathcal{A}_{\bar{q}}(V)$.

The polar decomposition of $S_{\Psi,\bar{q}}$ leads to the modular operator $\Delta^{\Psi}_{\bar{q}}$ such that \begin{equation}
\label{eq:modular_involution_sr}
    S_{\Psi,\bar{q}}=J_\Psi \Delta^{\Psi}_{\bar{q}}\,,
    \qquad
    \qquad
   \Delta^{\Psi}_{\bar{q}}=\Delta_\Psi\Pi_V(\bar{q}) \,.
\end{equation}
Notice that the modular conjugation $J_\Psi$ is the same in any charge sector and is equal to the one in the unresolved modular theory. The total modular operator thus decomposes into
\begin{equation}\label{eq:SRdec_modular_operator}
    \Delta_{\Psi}=\bigoplus_{\bar{q}}\Delta^{\Psi}_{\bar{q}}
    \,,
\end{equation}
and therefore $\Delta^{\Psi}_{\bar{q}}$ can be regarded as the \textit{symmetry-resolved modular operator} \cite{Di_Giulio_2023}. 
In the type I case, a similar relation to \eqref{eq:modopRDM0} also exists for the symmetry-resolved modular operator. If we describe the system via the state \eqref{eq:Schmidt_dec}, the reduced density matrix may be written as
\begin{equation}\label{eq:reduced_density_matrix_decomposition}
    \rho_V= \mathrm{tr}_{V'}\ket{\Psi}\!\bra{\Psi} =  \bigoplus_{\bar{q}} p(\bar{q}) \mathrm{tr}_{V'}\ket{\Psi_{\bar{q}}}\!\bra{\Psi_{\bar{q}}}\equiv \bigoplus_{\bar{q}} p(\bar{q})\rho_V(\bar{q}).
\end{equation}
We then write the symmetry-resolved modular operator as
\begin{equation}
\Delta^{\Psi}_{\bar{q}}=\rho_{V}(\bar{q})\otimes\rho_{V'}^{-1}(-\bar{q})\,.
\end{equation}
The charge $-\bar{q}$ in $V'$ is chosen such that $\Delta_{\bar{q}}^\Psi$ is represented on $\mathcal{H}_0$.
 Similarly to the unresolved case, where we can think of the modular operator as an extension of the reduced density matrix, we can think of the symmetry-resolved modular operator as an extension of the charge block $\rho_V(\bar{q})$.

Because of the decomposition of the algebra $\mathcal{A}(V)$ \eqref{eq:SRdec_algebra} and the modular operator \eqref{eq:SRdec_modular_operator}, we can also decompose the modular flow as \cite{Di_Giulio_2023}
\begin{equation}\label{eq:SR_modular_flow}
(\Delta_{\Psi})^{\mathrm{i}t}a(\Delta_{\Psi})^{-\mathrm{i}t}=\bigoplus_{\bar{q}}(\Delta^{\Psi}_{\bar{q}})^{\mathrm{i}t}a_{\bar q}(\Delta^{\Psi}_{\bar{q}})^{-\mathrm{i}t}\,,
\end{equation}
where $a=\bigoplus_{\bar{q}}a_{\bar q}\in\mathcal{A}(V)$ and $a_{\bar{q}}\in\mathcal{A}_{\bar{q}}(V)$. We call \eqref{eq:SR_modular_flow} \textit{symmetry-resolved modular flow}.
The symmetry-resolved modular flow naturally leads to the resolution of modular correlation functions defined as
\begin{equation}\label{eq:SR_modular_correlation_fct}
    G_{\textrm{\tiny mod}}(a_{\bar{q}},b_{\bar{q}};t,\bar q)= \langle\Psi_{{\bar q}}|b_{\bar q}\,(\Delta^{\Psi}_{\bar{q}})^{\mathrm{i}t}a_{\bar q}(\Delta^{\Psi}_{\bar{q}})^{-\mathrm{i}t}|\Psi_{\bar q}\rangle, \qquad
    a_{\bar{q}},b_{\bar{q}}\in\mathcal{A}_{\bar{q}}(V),
    \quad
    t\in\mathbb{R}\,,
\end{equation}
where, we have used the definition \eqref{eq:SR_Tomita_and_csstate}. From \eqref{eq:SR_Tomita_and_csstate}, \eqref{eq:SR_modular_flow}, and the idempotence of the projectors, we find \cite{Di_Giulio_2023}
\begin{equation}\label{eq:modular_correlation_function_decomposition}
    G_{\textrm{\tiny mod}}(a,b;t)=
    \sum_{\bar{q}}
    p(\bar q)G_{\textrm{\tiny mod}}(a_{\bar{q}},b_{\bar{q}};t,\bar q), \quad \forall a,b\in \mathcal{A}(V).
\end{equation}
It was shown in \cite{Di_Giulio_2023}, that the symmetry-resolved modular correlation function also satisfies the analyticity condition
\begin{equation}
\label{eq:resolved_KMS_AQFT}
 G_{\textrm{\tiny mod}}(a,b;t+\mathrm{i},\bar{q}) 
 =
\langle\Psi_{{\bar q}}|(\Delta^{\Psi}_{\bar{q}})^{\mathrm{i}t}a_{\bar q}(\Delta^{\Psi}_{\bar{q}})^{-\mathrm{i}t}\,b_{\bar q}|\Psi_{\bar q}\rangle\,.
\end{equation}
This is a remarkable result, as it shows that modular theory can be formulated consistently in each sector associated to a fixed subregion charge.

As mentioned above, in contrast to algebras of type II and type III, entanglement entropy is well-defined  and finite in type I systems.
In the setup discussed here, it is particularly useful to examine the entropy associated to the charge blocks,
\begin{equation}
    S_V(\bar{q})=-{\rm Tr}\left(\rho_{V}(\bar{q})\ln\rho_{V}(\bar{q}) \right)\,, 
    \label{eq:def_SRentanglemententropy}
\end{equation}
which is called the \textit{symmetry-resolved entanglement entropy}. The full entanglement entropy associated with $V$ is then
\begin{equation}\label{eq:full_entanglement_entropy_decomp}
    S_V=-{\rm Tr}\left(\rho_{V}\ln\rho_{V}\right)=
    \sum_{\bar{q}}\left[p(\bar{q})S_V(\bar{q})-p(\bar{q})\ln p(\bar{q})\right]\,.
\end{equation}
Here, the first term  is referred to as configurational entropy, whereas the second term is referred to as fluctuation entropy. These terms encode information about entanglement within each charge sector, and the entropy in the distribution $p(q)$, respectively.

The discussion so far relied  on the decomposition of the Hilbert space as given by \eqref{eq:chargesectors}, which however does not hold for general QFTs, due to the  short wavelength modes across the interface of $V$ and $V'$. Algebraically speaking this means that the above discussion is not valid for the hyperfinite algebras that include type II and type III algebras. The goal of this work is to extend this analysis to these cases. In the following we will use a regularized version of the entanglement entropy and its resolution to give a physical intuition for our results.

\subsection{Review of the infinite tensor product construction}\label{sec:AW_setup}

\begin{table}
    \centering
    \begin{tabular}{c|c|c|c}
         \textbf{Type} & \textbf{\makecell{Finite entanglement \\ state?}} & \textbf{Trace existent?} & \textbf{\makecell{Accumulation points of \\
         spectrum of $\Delta_\Psi$}} \\\hline\hline
     I$_d$ & yes & yes & discrete, positive\\\hline
     I$_\infty$ &yes & for particular operators & discrete, positive\\\hline\hline
     II$_1$ & no & yes & $\{1\}$\\\hline
     II$_\infty$ & no & for particular operators & $\{1\}$\\\hline\hline
     III$_\lambda$ & no & no & $\{0,\lambda^k\vert k\in\mathbb{Z}\}$\\\hline
     III$_0$ & no & no & $\{0,1\}$\\\hline
     III$_1$ & no & no & $[0,\infty)$
    \end{tabular}
    \caption{Classification of von Neumann algebras. }
    \label{tab:vn_class}
\end{table}

In contrast to type I, in type II and type III algebras, the spatial bipartition property \eqref{eq:hilbert_spatial_decomp} holds no longer true. Algebras of type II and type III do not act irreducibly on the Hilbert space. In type II algebras, however, there exists a trace functional, i.e.~a cyclic linear functional, and thus, for some operators, it is still possible define an entanglement entropy. The existence of the trace is linked to the existence of a state with maximal, formally infinite, entanglement in the Hilbert space. There are two subtypes -- II$_1$, where the trace functional is defined for all operators in the algebra, and II$_\infty$, where the trace is finite only for a subset of all operators. Type III algebras do not permit a trace and it is thus impossible to define reduced density matrices. They can be further classified according to the spectrum of the corresponding modular operator. For type III$_\lambda$ the spectrum has a countable number of accumulation points corresponding to integer powers of $\lambda$. For type III$_0$ the spectrum of the modular operator has accumulation points $0$ and $1$, whereas for type III$_1$, the spectrum consists of the non-negative reals, $[0,\infty)$. The resulting classification is summarized in table \ref{tab:vn_class}.

In order to extend the analysis of \secref{sec:modular_theory} and \secref{sec:type_I_resolution} to type II and type III algebras, we use a construction introduced in \cite{Powers1967,Araki1968} consisting of building these algebras from infinite tensor products of finite-dimensional ones. We denote the algebras constructed in this way as infinite tensor product factors of type I (ITPFI), following the traditional nomenclature \cite{Takesaki1979}. This construction is useful since it allows for the description of type II$_1$, III$_\lambda$, III$_1$ and III$_0$ algebras in a single framework by tuning a sequence of parameters. Furthermore, all algebras with a trivial centre, called factor algebras, of the respective subtypes are isomorphic to the corresponding ITPFI. This makes the analyses valid for a broad range of algebras, despite the precise isomorphism relating the physical algebras to the ITPFIs being potentially highly non-trivial. A more extensive introduction to this construction can be found e.g.~in \cite{Witten:2018zxz}.

\begin{figure}
    \centering
    \def\svgwidth{0.3\columnwidth}
    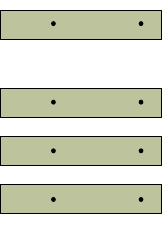
    \caption{Setup of the ITPFI construction. The Hilbert space is constructed from an infinite tower of pairwise entangled spins.}
    \label{fig:aw_setup}
\end{figure}

To construct these algebras, we consider two infinite towers 
of pairwisely entangled spins as shown in \figref{fig:aw_setup}.
The Hilbert space of each spin pair is isomorphic to the space of complex $2\times2$ matrices. For physical intuition, we can identify the basis elements with tensor products of spin states,
\begin{equation}\label{eq:basis_identification}
    \ket{\uparrow}_L\ket{\downarrow}_R\leftrightarrow \begin{pmatrix}
        1 & 0\\
        0 & 0
    \end{pmatrix}\,,\quad \ket{\uparrow}_L\ket{\uparrow}_R\leftrightarrow\begin{pmatrix}
        0 & 1\\
        0 & 0
    \end{pmatrix}\,, \quad \ket{\downarrow}_L\ket{\downarrow}_R\leftrightarrow\begin{pmatrix}
        0 & 0\\
        1 & 0
    \end{pmatrix}\,, \quad \ket{\downarrow}_L\ket{\uparrow}_R\leftrightarrow\begin{pmatrix}
        0 & 0\\
        0 & 1
    \end{pmatrix}\,.
\end{equation}
We choose this basis such that the diagonal elements correspond to physical configurations for which the sum of the eigenvalues of $\sigma_z$ of the two spins vanishes.

One would naively expect the Hilbert space of the full spin system to be the infinite tensor product of $2\times2$ matrices. This results, however, in a Hilbert space of uncountably infinite dimension. To avoid this, consider the state
\begin{equation}
        \label{eq:cyclicseparating AW}
        \ket{\Psi_{\boldsymbol{\lambda}}}=K_{2,\lambda_1}\otimes K_{2,\lambda_2}\otimes K_{2,\lambda_3}\otimes \dots\otimes K_{2,\lambda_i}\otimes\dots\,.
\end{equation}
Here,
\begin{equation}
        \label{eq:Kmatrix AW}
        K_{2,\lambda_i}=\frac{1}{\sqrt{1+\lambda_i}}\begin{pmatrix}
             1& 0 \\
             0& \sqrt{\lambda_i}
        \end{pmatrix}\leftrightarrow \frac{1}{\sqrt{1+\lambda_i}}\left(\ket{\uparrow}_L\ket{\downarrow}_R+ \sqrt{\lambda_i} \ket{\downarrow}_L\ket{\uparrow}_R\right)\,,
\end{equation}
 where $0<\lambda_i<1$. From now on, we use the label $\boldsymbol{\lambda}$ as a shorthand notation for the $\lambda_i$. The state entangles the left and right spins in such a way that the entanglement is maximal for $\lambda_i=1$ and zero for $\lambda_i=0$. The state $\ket{\Psi}$ will play the role of the cyclic separating state.

We now define the restricted tensor product pre-Hilbert space
\begin{equation}
\label{eq:prehilbspace}
    \tilde{\mathcal{H}}=\{\ket{\theta}=\theta_1\otimes\dots\otimes \theta_k\otimes K_{2,\lambda_{k+1}}\otimes \dots\}\,,
\end{equation}
where the $\theta_i$ are arbitrary $2\times 2$ complex matrices. $\tilde{\mathcal{H}}$ consists of an infinite tensor product where all but $k$ factors are the matrices $K_{2,\lambda_i}$. Crucially, the number $k$ can be arbitrarily large. The dimension of $\tilde{\mathcal{H}}$, while still infinite, is thus countable \cite{Witten:2018zxz}.
The pre-Hilbert space $\tilde{\mathcal{H}}$ admits a Hermitian inner product. Consider the states $\ket{\theta}=\theta_1\otimes\dots\otimes \theta_k\otimes K_{2,\lambda_{k+1}}\otimes\dots$ and $\ket{\phi}=\phi_1\otimes\dots\otimes \phi_n\otimes K_{2,\lambda_{n+1}}\otimes\dots$,  where without restricting generality, we assume $k>n$. We define
\begin{equation}\label{eq:inner_product_hilbert_space}
    \braket{\theta}{\phi}=\prod_{i=1}^k \mathrm{tr}(\theta_i^\dagger \phi_i)\,,
\end{equation}
where the trace is the trace of the $2\times2$ matrices.
The completion of $\tilde{\mathcal{H}}$ with respect to this inner product is called the restricted tensor product Hilbert space,
\begin{equation}\label{eq:AW_Hilbert_space_definition}
    \mathcal{H}=\overline{\tilde{\mathcal{H}}}=\overline{\{\ket{\theta}=\theta_1\otimes\dots\otimes \theta_k\otimes K_{2,\lambda_{k+1}}\otimes \dots\}}\,.
\end{equation}
A similar procedure can also be applied to obtain the algebra of bounded operators acting on $\mathcal{H}$. Each spin pair admits two commuting algebras of operators, namely the space of $2\times 2$ complex matrices acting by multiplication from the left and right respectively. We interpret the left algebra as corresponding to subregion $V$ and the right algebra as corresponding to the complement $V'$ and denote the respective algebras by $L$ and $R$. As a starting point, consider the algebra
\begin{equation}
        \label{eq:operatorAW}
        \tilde{\mathcal{F}}(L)=\{a=a_1\otimes a_2\otimes\dots\otimes a_k\otimes\1\otimes \1\otimes\dots\}\,,
\end{equation}
where $\1$ is the $2\times2$ unit matrix and $a_i$ are arbitrary complex $2\times2$ matrices that act on the states by multiplication from the left. This algebra preserves $\tilde{\mathcal{H}}$. In order to obtain the full algebra acting on $\mathcal{H}$, consider the closure of $\tilde{\mathcal{F}}(L)$ by including limits of sequences $a^{(n)}\in\tilde{\mathcal{F}}(L)$ such that $a^{(n)}\ket{\theta}\in\mathcal{H}$ converges. In fact, there are two ways to make sense of this statement, called \textit{weak} and \textit{strong convergence}, respectively. An operator is said to converge weakly if for all $\ket{\theta},\ket{\phi}\in \mathcal{H}$, $\braket{\phi}{a^{(n)}\theta}$ converges. That is, the sequence of $n$ needs to be a Cauchy sequence, i.e.~$\forall\epsilon>0$ there exists $k$ such that for all $n,m>k$
\begin{equation}\label{eq:cauchy_seq_condition}
    \left|\braket{\phi}{\left(a^{(n)}-a^{(m)}\right)\theta}\right|<\epsilon\,.
\end{equation}
In contrast, a sequence of operators is said to converge strongly, if for all $\ket{\phi}\in\mathcal{H}$ and for all $\epsilon>0$ there exists $k$ such that for all $n,m>k$,
\begin{equation}\label{eq:strong_convence_criterion}
    ||(a^{(n)}-a^{(m)})\ket{\phi}||<\epsilon.
\end{equation}
Here $||\cdot||$ is any norm, e.g.~the one induced by the inner product \eqref{eq:inner_product_hilbert_space}.
In fact, in a von Neumann algebra as in the present construction, weak and strong convergence are equivalent, i.e.~any weakly convergent sequence of operators is also strongly convergent and vice versa.

We can thus reasonably define the action of the limit of the sequence by
\begin{equation}
    a \ket{\theta}\equiv\ket{a\theta}=\lim_{n\to\infty} a^{(n)}\ket{\theta}\,,
\end{equation}
where the matrix elements are defined by
\begin{equation}\label{eq:convergence_matrix_elements}
    \braket{\phi}{a\theta}=\lim_{n\to\infty} \braket{\phi} {a^{(n)}\theta}\,.
\end{equation}
Upon including such limits, we obtain the algebra
\begin{equation}
\label{eq:subregion_algebra_itpfi}
    \mathcal{F}(L)=\overline{\tilde{\mathcal{F}}(L)}=\overline{\{a=a_1\otimes a_2\otimes\dots\otimes a_k\otimes\1\otimes \1\otimes\dots\}}\,.
\end{equation}
The resulting von Neumann algebra $\mathcal{F}(L)$ is therefore characterized by the properties of the cyclic and separating vector (\ref{eq:cyclicseparating AW}) and therefore of the matrix (\ref{eq:Kmatrix AW}). More precisely, the values for $\lambda_i$ determine the type of the algebra; for $\lambda_i=1$, we find a type II$_1$ algebra, for $\lambda_i=\lambda^*$ we find a type III$_\lambda$ algebra, for $\lambda_i\to0$, we find type III$_0$ and for general $\lambda_i$ we obtain a type III$_1$ algebra \cite{Witten:2018zxz}. In other words, the type of the algebra is here not determined by the form of the operators but on the way in which such operators act in $\mathcal{H}$.
Repeating the same procedure, we obtain the algebra $\mathcal{F}(R)$. Since $\mathcal{F}(R)$ is the commutant of $\mathcal{F}(L)$ we can then define
\begin{equation}
    \mathcal{F}=\mathcal{F}(L)\otimes \mathcal{F}(R)\,.
\end{equation}
This algebra thus contains operators that act on both the left and right tower of spins.

\section{Conserved charges in infinite tensor product algebras}
\label{sec:firstapproach}


Based on the construction reviewed in \secref{sec:AW_setup}, we now establish how to define a conserved charge and its restriction to local subregions in ITPFI. Our proposed construction for symmetry resolution based on these considerations is presented below in \secref{sec:hyeprfiniteresolved}.

\subsection{Local magnetization operators}
\label{subsec:chargenorescaling}

In contrast to the type I case described in \secref{sec:type_I_resolution}, in hyperfinite algebras the subregion charge operator has an infinite number of non-trivial tensor product factors. This implies that it is no longer an element of $\mathcal{\tilde{F}}(L)$ in \eqref{eq:operatorAW}. In other words, the charge will be the limit of a convergent series of operators with respect to the inner product \eqref{eq:inner_product_hilbert_space}. In analogy to QFT, we consider subregion charge operators that have access to an arbitrary number of modes localized near the boundary of the subsystems, instead of a UV cutoff version of the subregion charge.


To keep with the intuition of the infinite tensor product as a tower of entangled spins, we consider {\it local magnetization operators} of the individual spins in the left and right subsystems and their sums. Thus, similarly to \eqref{eq:operatorAW}, we can consider operators of the form
\begin{equation}
   z_L^i=\underset{i-1~\text{times}}{\1\otimes...\otimes\1}\otimes\underset{i\text{-th spin}}{\frac{\sigma_z}{2}}\otimes\1\otimes\dots \in \mathcal{F}(L)\,,
\end{equation}
where $\sigma_z$ is the third Pauli matrix. We may think of this operator as the magnetization operator of the $i$-th spin in the left tower. In fact, note that $z_L^i\in \tilde{\mathcal{F}}(L)$ for any $i$, i.e.~the non-trivial part of the product truncates after the $i$-th spin. This does not apply to a generic operator in $\mathcal{F}(L)$, which may not belong to $\tilde{\mathcal{F}}(L)$. We then define the magnetization operator for the first $N$ spins in the left tower,
\begin{equation}\label{eq:subregion_magnetization_sequence_defintion}
    Z_L^{(N)}=\sum_{i=1}^N z_L^i\,.
\end{equation}
This operator is an element of $\tilde{\mathcal{F}}(L)$ for arbitrary $N$. We similarly define the local magnetization for the $i$-th spin in the right tower,
\begin{equation}
    z_R^i=\underset{i-1~\text{times}}{\1\otimes...\otimes\1}\otimes\underset{i\text{-th spin}}{\frac{-\sigma_z}{2}}\otimes\1\otimes\dots \in \mathcal{F}(R)\,,
\end{equation}
where the negative sign stems from the identification of the basis \eqref{eq:basis_identification}, where the left and right spin operators were defined with opposite signs along the diagonal.
We can then also define $Z_R^{(N)}$ analogously as for the left subsystem by summing $z_R^i$ for the first $N$ spins. Furthermore, we define
\begin{equation}\label{eq:full_mag_finite_N}
    Z^{(N)}=Z_L^{(N)}\oplus Z_R^{(N)}\in \mathcal{F}\,,
\end{equation}
where the direct sum is well defined as $\mathrm{id}_L=\1\otimes\1\otimes\dots\in \mathcal{F}(L)$ and $\mathrm{id}_R\in \mathcal{F}(R)$ analogously, i.e.~$\mathcal{F}(L)$ and $\mathcal{F}(R)$ are unital algebras. The cyclic separating state \eqref{eq:cyclicseparating AW}, by construction, is an eigenstate of $Z^{(N)}$ with eigenvalue $0$,
\begin{equation}\label{eq:eigenstate_AW}
    Z^{(N)}\ket{\Psi_{\boldsymbol{\lambda}}}=0\,.
\end{equation}
This can be checked by noting that the matrix \eqref{eq:Kmatrix AW} commutes with $\sigma_z$ and the sum of left and right magnetization of each spin pair vanishes. This mirrors \eqref{eq:eigenstatecharge} and, in analogy with the nomenclature of \secref{sec:modular_theory}, we refer to $Z^{(N)}$ as charge.
Consequently, $Z_L^{(N)}$ and $Z_R^{(N)}$ play the role of subregion charges in the left and right subsystem, respectively. 

The spectrum of $Z_L^{(N)}$ is given by
\begin{equation}\label{eq:mag_finite_N_spectrum}
    {\rm Spec}(Z_L^{(N)})=\left\{-\frac{N}{2}, -\frac{N}{2}+\frac{1}{2},\dots, \frac{N}{2}-\frac{1}{2}, \frac{N}{2}\right\}\,,
\end{equation}
where each eigenvalue labels a charge sector. Note that the eigenvalues are either integer or half-integer valued for $N$ even or odd, respectively. 
If we consider the symmetry generated by exponentiating $Z^{(N)}$, in the limit $N\to \infty$, we obtain a U(1) symmetry.
We may also consider bounded functions of the operators defined above, for instance for $m\in \sigma(Z_L^{(N)})$,  \cite{Xavier:2018kqb}
\begin{equation}\label{eq:projectors_finite_N}
        \Pi_L^{(N)}(m)=\frac{1}{N+1}\sum_{n=-\frac{N}{2}}^{\frac{N}{2}}\exp\left(\frac{2\pi i n}{N+1}(Z_L^{(N)}-m\;\mathrm{id}_L)\right)\,.
\end{equation}
This operator projects to the $m$-th eigenspace of $Z_L^{(N)}$, i.e.~to the charge sector labeled by $m$.

In the limit $N\to \infty$, the operators defined in \eqref{eq:projectors_finite_N}, \eqref{eq:full_mag_finite_N}, and \eqref{eq:subregion_magnetization_sequence_defintion}, act on an infinite number of spin pairs in the tower. In fact, only the total magnetization operator \eqref{eq:full_mag_finite_N} converges and is an element of $\mathcal{F}$. We provide a detailed discussion of this fact in \appref{app:convergence_ZN_ZNL_ZNR}. In order to have a local magnetization operator that is well defined in the $N\to\infty$ limit, we have to modify the definition \eqref{eq:subregion_magnetization_sequence_defintion}, which is done in the next subsection.

\subsection{Rescaling of the charge}
\label{subsec:rescaling}


To motivate the modification of the magnetization operator given in \eqref{eq:subregion_magnetization_sequence_defintion}, we follow the logic of symmetry-resolved entanglement entropy \eqref{eq:def_SRentanglemententropy} in QFT, which consists of rescaling the charge by the UV regulator. Symmetry-resolved entanglement has been studied in two-dimensional CFTs with U(1) global symmetries for several choices of subsystems $V$ \cite{Goldstein:2017bua,Xavier:2018kqb, Ares:2022gjb,DiGiulio:2022jjd}. All of these examples involve a UV cutoff $\varepsilon$. 

The need for a rescaling of the charge arises as follows. The introduction of the cutoff amounts to choosing a local operator algebra in the  subregion $V$ of type I.  For $V$ an interval of length $\ell$, it was found that at leading order in the UV cutoff, the probability distribution \eqref{eq:prob_def} reads \cite{Xavier:2018kqb}
\begin{equation}\label{eq:probability_CFT}
    p(\bar{q})\simeq\sqrt{\frac{\pi}{\ln(\frac{\ell}{\varepsilon})}}e^{-\frac{\pi^2 \bar{q}^2}{\ln(\frac{\ell}{\varepsilon})}}\,.
\end{equation}
To retrieve the case of an hyperfinite algebra in the subregion $V$, the UV cutoff has to be removed, i.e.~$\varepsilon\to0$. Removing the cutoff results in a pathological probability distribution. Indeed, \eqref{eq:probability_CFT} 
pointwisely converges to a flat distribution identically equal to $0$. Crucially, this assumes that the charge is independent of the cutoff. To resolve this conundrum, one takes the charge $\bar{q}$ itself to scale with $\sqrt{\ln(\ell/\varepsilon)}$. This implies that, as $\varepsilon$ gets smaller, charges away from the peak of the distribution ($\bar{q}=0$ in this case) get suppressed and the analysis becomes reliable only in the vicinity of the peak. We define a new, rescaled, subregion charge variable
\begin{equation}\label{eq:sr_rescaled_charge}
    \tilde{q}=\frac{\pi\, \bar{q}}{\sqrt{\ln(\frac{\ell}{\varepsilon})}}\,.
\end{equation}
Even if $\bar{q}$ is discrete, in the limit $\varepsilon\to 0$, $\tilde{q}$ will take continuous values, since the spacing between consecutive charge values decreases to $0$. Changing variables and taking into account the Jacobian to preserve the normalization, the distribution \eqref{eq:probability_CFT} becomes
\begin{equation}
    \label{eq:probability_CFT_cont}p(\tilde{q})\simeq\frac{1}{\sqrt{\pi}}e^{-\tilde{q}^2}\,.
\end{equation}
In the CFT example above, the rescaling makes the charge probability distribution in a subregion well defined and non-trivial, at the price of rendering the values taken by the charge continuous.
Inspired by this analysis, to introduce a consistent subregion charge in our setup, we define the rescaled left charge
\begin{equation}\label{eq:rescaled_subregion_charge_def}
    s_L^{(N)}=\frac{1}{N} Z_L^{(N)}=\frac{1}{N}\sum_{i=1}^N z_L^i\,.
\end{equation}
The same can be done for the right charge. The parameter $N$ in this case plays the role similar to the cutoff in \eqref{eq:sr_rescaled_charge}. From \eqref{eq:mag_finite_N_spectrum}, we can infer the spectrum of $s_L^{(N)}$, consisting of numbers from $-\frac{1}{2}$ to $\frac{1}{2}$ with spacing $\frac{1}{2N}$. As $N\to\infty$, the spacing goes to $0$, and the spectrum becomes continuous in the interval $[-1/2,1/2]$. This is consistent with the interpretation of $s_L^{(N)}$ as being the average magnetization per spin (or magnetization density) in the left tower.

To ensure that the rescaled operator \eqref{eq:rescaled_subregion_charge_def} can be indeed a meaningful charge for our purposes, we examine whether it converges according to \eqref{eq:strong_convence_criterion} in the $N\to\infty$ limit.
Intuitively, we can understand the limit in the following way: The expectation value in a generic state in $\mathcal{\tilde{H}}$ of the operator $s_L^{(N)}$ will be dominated by the contributions of the matrices \eqref{eq:Kmatrix AW} that constitute the cyclic separating vector as $N$ becomes large. Even in the full Hilbert space, the tensor product factors of the states rapidly approach the matrices \eqref{eq:Kmatrix AW}, and the same argument applies. As the expectation value depends only on these dominant contributions for all the states in $\mathcal{{H}}$, the operator $s_L^{(N)}$ behaves effectively in the same way with respect to any state as $N$ is large. We thus conjecture
\begin{equation}
\label{eq:guess_identity}
    \lim_{N\to\infty} s_L^{(N)}\propto \mathrm{id}_L\,.
\end{equation}
In QFT language, the operator $s_L^{(N)}$ is dominated by modes of very short wavelength and any vector at short distances can be approximated well by the vacuum.
In fact, the prefactor in \eqref{eq:guess_identity} should match the average magnetization per spin in the left tower in the cyclic separating state \eqref{eq:cyclicseparating AW}, which is given by $\frac{1}{2}\frac{1-\lambda}{1+\lambda}$ for algebras of type III$_\lambda$ and $0$ for type II$_1$. 
Now, we are ready to check the conjecture \eqref{eq:guess_identity} for all the possible choices of $\boldsymbol{\lambda}$, i.e.~for different types of algebras.

\paragraph{Type III$_\lambda$:} We are thus ready to examine whether
\begin{equation}\label{eq:rescaled_charge_typeIII_lambda_limit}
    s_L\equiv \lim_{N\to\infty} s_L^{(N)} = \frac{1}{2}\frac{1-\lambda}{1+\lambda}\mathrm{id}_L\,.
\end{equation}
To prove this statement, it is enough to check that
\begin{equation}
    \left|\left|\left(s_L^{(N)}-\frac{1}{2}\frac{1-\lambda}{1+\lambda}\right) \ket{\theta} \right|\right|\leq \mathcal{O}(N^{-1})\,,
\end{equation}
for arbitrary $\ket{\theta}\in\tilde{\mathcal{H}}$. This inequality can be proven by an explicit calculation. Since the r.h.s.~tends to $0$ as $N\to\infty$, we conclude that $s_L^{(N)}$ indeed converges on $\tilde{\mathcal{H}}$. Since $\tilde{\mathcal{H}}$ is dense in $\mathcal{H}$, the limit conjectured in \eqref{eq:rescaled_charge_typeIII_lambda_limit} is correct.

\paragraph{Type II$_1$ and III$_0$:} These cases correspond to special cases of the above analysis when we set $\lambda=1$ and $\lambda=0$ respectively.

\paragraph{Type III$_1$:} Consider the cyclic separating vector \eqref{eq:cyclicseparating AW} and the expectation value
\begin{equation}
\label{eq:III_1}
    \braket{\Psi_{\boldsymbol{\lambda}}}{s_L^{(N)}\Psi_{\boldsymbol{\lambda}}}=\frac{1}{2}\frac{1}{N}\sum_{i=1}^N \frac{1-\lambda_i}{1+\lambda_i}\,.
\end{equation}
The convergence of the r.h.s.~is non-trivial and, therefore, we need a careful examination.
In general, it will not converge for a (non-convergent) sequence $\{\lambda_i\}$. An example of the r.h.s.~of \eqref{eq:III_1} not converging is given in \appref{app:convergence_rescaled_carge_III_1}. 
However, there are instances where $s_L^{(N)}$ converges (see \appref{app:convergence_rescaled_carge_III_1}). In this case it still converges to a multiple of the identity. To show this, we consider an operator of the form
\begin{equation}\label{eq:general_rescaled_operator}
    a^{(N)}=\frac{1}{N}\sum_{i=1}^N a_i \in\mathcal{F}(L)\,,
\end{equation}
where
\begin{equation}
    a_i=\underbrace{\1\otimes\dots\otimes}_{i-1\text{ times}} a \otimes\1\otimes\dots\,.
\end{equation}
Operators of this form converge, when $N\to\infty$, to multiples of the identity if at all. Consider $\ket{\theta},\ket{\varphi}\in\tilde{\mathcal{H}}$, such that $p$ denotes the number of non-trivial tensor product factors in \eqref{eq:prehilbspace}. Without loss of generality, we assume $N>p$. Then
\begin{equation}
    \braket{\theta}{a^{(N)}\varphi}=\frac{1}{N}\sum_{i=1}^p \braket{\theta}{a_i\varphi}+\frac{\braket{\theta}{\varphi}}{N}\sum_{i=p+1}^N\braket{\Psi_{\boldsymbol{\lambda}}}{a_i\Psi_{\boldsymbol{\lambda}}}.
\end{equation}
As $N\to\infty$, the first term on the right hand side will be suppressed and the matrix element will be dominated by the second term which only depends on the inner product of $\ket{\theta},\ket{\varphi}$. Thus, if the limit
\begin{equation}
    \lim_{N\to\infty} \frac{1}{N}\sum_{i=p+1}^N \braket{\Psi_{\boldsymbol{\lambda}}}{a_i\Psi_{\boldsymbol{\lambda}}}
\end{equation}
exists, also the limit of the sequence $a^{(N)}$ will exists but it will be proportional to $\mathrm{id}_L$ on $\tilde{\mathcal{H}}$. Since $\tilde{\mathcal{H}}$ is dense in $\mathcal{H}$, it will also converge to the same operator on $\mathcal{H}$.
Comparing with \eqref{eq:rescaled_subregion_charge_def}, we observe that $s_L^{(N)}$ is of the form \eqref{eq:general_rescaled_operator}, and, therefore 
\begin{equation}\label{eq:rescaled_charge_type_III1}
    s_L\equiv \lim_{N\to\infty}s_L^{(N)}=\frac{1}{2}\lim_{N\to\infty}\frac{1}{N}\sum_{i=1}^N\frac{1-\lambda_i}{1+\lambda_i} \mathrm{id}_L\,.
\end{equation}
This equation implies that the local algebras are characterized by a single charge sector associated with $s_L$. 

In the following discussion, we  denote the proportionality factor of $s_L$, as defined in \eqref{eq:rescaled_charge_typeIII_lambda_limit} and \eqref{eq:rescaled_charge_type_III1}, by $\bar{q}$. In doing so, we anticipate the discussion of \secref{sec:hyeprfiniteresolved} where we will combine algebras with different proportionality factors to obtain a larger algebra with non-trivial centre. By rescaling the subregion charge magnetization, we have constructed (at least in the case of type II$_1$, III$_0$ and III$_\lambda$ and in some instances of type III$_1$) an operator that has access to all tensor product factors. 
In \secref{sec:hyeprfiniteresolved}, we will exploit this result to explicitly build the algebra of invariant operators that we are interested in. We will then use the resulting algebra and its decomposition to study the symmetry resolution of modular operator and modular flow defined for hyperfinite algebras.


\subsection{Interpretation through symmetry-resolved entanglement}

To understand the emergence of a single charge sector in the large $N$ limit, as explained in \secref{subsec:rescaling}, we use symmetry-resolved entanglement entropy as defined in \eqref{eq:def_SRentanglemententropy}. However, there is the complication that entanglement entropies are divergent in hyperfinite algebras.
To circumvent this problem, we introduce a regularized entanglement entropy and study its leading term in the $N\to\infty$ limit.
This procedure is physically intuitive but mathematically less rigorous than the calculations presented in \secref{subsec:rescaling} and should be viewed as complementary evidence to support its conclusion.

More precisely, we consider the entanglement entropy in the state
\begin{equation}
    \ket{\Psi_{N,\boldsymbol{\lambda}}}=K_{2,\lambda_1}\otimes\dots\otimes K_{2,\lambda_N}\,,
\end{equation}
and its symmetry resolution in the sectors of the symmetry generated by the corresponding subregion magnetization operator
and consider the divergent leading order behaviour as $N\to\infty$. Since the Hilbert space for finite $N$ has dimension $4^N<\infty$, entanglement entropies are well defined. The starting point for analysing entanglement entropy is the reduced density matrix, $\rho_L^{(N)}=\mathrm{Tr}_R(\ket{\Psi_{N,\boldsymbol{\lambda}}}\bra{\Psi_{N,\boldsymbol{\lambda}}})$, where $\mathrm{Tr}_R$ is the partial trace as defined in quantum mechanics. In order to calculate $\rho_L^{(N)}$, it is advantageous to use the analogy to spins presented on the right side of \eqref{eq:Kmatrix AW}. In fact, we can label the states by their total subregion magnetization. Of course, there are many combinations of $N$ spins which form a state with fixed magnetization, so the states will be labelled by $\ket{m,i}$, where $m$ is the eigenvalue of the subregion magnetization operator and $i$ labels the degeneracy. For given $N$, $m$ takes values $\{-\frac{N}{2},-\frac{N}{2}+\frac{1}{2},\dots, \frac{N}{2}\}$. There are $\binom{N}{\frac{N}{2}-m}$ possible states for a given value of $m$, and thus $i\in\{1,\dots,\binom{N}{\frac{N}{2}-m}\}$. 
In this basis, the reduced density matrix reads
\begin{equation}\label{eq:finite_N_density_matrix}
    \rho_L^{(N)}
=\frac{1}{\prod_{i=1}^N(1+\lambda_i)}\sum_{m=-\frac{N}{2}}^{\frac{N}{2}}\sum_{i=1}^{\binom{N}{\frac{N}{2}-m}}\text{Pol}_{\frac{N}{2}-m,i}(\boldsymbol{\lambda})\ket{m,i}\bra{m,i}_N^L\,.
\end{equation}
Here, $\text{Pol}_{\frac{N}{2}-m,i}(\boldsymbol{\lambda})$ is a polynomial of the $\lambda_i$ of degree $\frac{N}{2}-m$. 
The expressions of these polynomials can be derived with a long and straightforward computation, but those expressions are not directly relevant for our analysis.
From now on, in this subsection, 
we drop for convenience the range of the indices in the two sums over $m$ and $i$.
Viewed as a sequence of operators, the reduced density matrices do not converge as $N\to\infty$, as expected, since the (quantum mechanical) partial trace is ill-defined in hyperfinite algebras. However, at finite $N$ the reduced density matrix is well-defined. A similar expression can be derived for the reduced density matrix of the right subsystem. We can calculate the entanglement entropy associated to $\rho_L^{(N)}$ by taking its von Neumann entropy.
We are interested in the type II$_1$, III$_\lambda$ and III$_0$ case, which are accessed by choosing $\lambda_i=\lambda$ and subsequently taking the limit $N\to\infty$. Setting $\lambda_i=\lambda$, the polynomials in \eqref{eq:finite_N_density_matrix} become $\lambda^{\frac{N}{2}-m}$, since all $\lambda_i$ are equal, and we obtain the regularized entanglement entropy
\begin{eqnarray}
    S_L^{(N)}&=&\frac{1}{(1+\lambda)^N}\sum_m \binom{N}{\frac{N}{2}-m}\lambda^{\frac{N}{2}-m}\log\left(\frac{\lambda^{\frac{N}{2}-m}}{(1+\lambda)^N}\right)
    \nonumber
    \\
\label{eq:full_entanglement_entropy_finite_N}
    &=& -N\left(-\frac{1}{2}\frac{1-\lambda}{1+\lambda}\log{\lambda}+\log{\frac{\sqrt{\lambda}}{1+\lambda}}\right)\,,
\end{eqnarray}
where the superscript explicitly indicates that we are considering the entropy regularized by $N$. 
Indeed, unless $\lambda=0$ when there is no entanglement between the degrees of freedom, $S_L^{(N)}$ clearly diverges linearly as $N\to\infty$.

Now, we turn to the symmetry-resolved entanglement entropy. Comparing \eqref{eq:finite_N_density_matrix} to \eqref{eq:reduced_density_matrix_decomposition}, we can read off the reduced density matrices in each charge sector.
When $\lambda_i=\lambda$, we obtain
\begin{equation}
   \rho_L^{(N)}(m) = \frac{\lambda^{\frac{N}{2}-m}}{(1+\lambda)^N}\frac{1}{p_m^{(N)}} \sum_{i} \ket{m,i}\bra{m,i}_N^L\label{eq:charge_sector_finite_N_unrescaled}\,.
\end{equation}
Moreover, for the same parameters, the probability distribution \eqref{eq:prob_def} associated with the unrescaled magnetization reads
\begin{align}
     p_m^{(N)}=\frac{1}{(1+\lambda)^N}\binom{N}{\frac{N}{2}-m}\lambda^{\frac{N}{2}-m}\label{eq:prob_unrescaled_finite_N}\,.
\end{align}
This probability distribution is equivalently described by a binomial distribution with elementary probability $\frac{\lambda}{\lambda+1}$, mean $\frac{N}{2}\left(\frac{1-\lambda}{1+\lambda}\right)$ and variance $\frac{N\lambda}{(1+\lambda)^2}$. In the limit $N\to\infty$, the variance of the distribution diverges and the probability distribution converges to $0$. This analysis supports the conclusion of the above subsection, i.e.~the unrescaled magnetization is not a useful quantity to consider in the $N\to \infty$ limit. Instead, we should examine the probability distribution of the variable $\bar{q}=\frac{m}{N}$. The charge $\bar{q}$ takes values in the interval $[-\frac{1}{2},\frac{1}{2}]$ in increments of $\frac{1}{N}$. The distribution of the variable $\bar{q}$ is described at finite $N$ by a binomial distribution with mean $\frac{1}{2}\left(\frac{1-\lambda}{1+\lambda}\right)$ and variance $\frac{\lambda}{N(1+\lambda)^2}$. In considering the rescaled variable, in the $N\to\infty$ limit, the variable $\bar{q}$ takes values close to any real number in the interval $[-\frac{1}{2},\frac{1}{2}]$, as the increment tends to $0$, and its probability distribution is approximately described by a continuous distribution.
Moreover, in this limit, the variance tends to zero while the mean stays constant. Thus, the probability distribution of the rescaled magnetization, when defined in an appropriate way, in the limit $N\to\infty$ is given by $\delta(\bar{q}-\frac{1}{2}\frac{1-\lambda}{1+\lambda})$.
The delta-distribution indicates the presence of a single charge sector. The position of the peak precisely coincides with the prefactor of the rescaled charge operator in \eqref{eq:rescaled_charge_typeIII_lambda_limit}, supporting our findings in \secref{subsec:rescaling}.

Finally, we calculate the symmetry-resolved entanglement entropy as defined in \eqref{eq:def_SRentanglemententropy} using the expression of the charge block in \eqref{eq:charge_sector_finite_N_unrescaled}. We obtain
\begin{equation}
    S_L^{(N)}(m)=\log\binom{N}{\frac{N}{2}-m}\,.
\end{equation}
Together with the probability distribution \eqref{eq:prob_unrescaled_finite_N}, the decomposition \eqref{eq:full_entanglement_entropy_decomp} of the full entanglement entropy \eqref{eq:full_entanglement_entropy_finite_N} can be explicitly verified. Using the rescaled charge $\bar{q}=\frac{m}{N}$, we expand the symmetry-resolved entanglement entropy to leading order in $N$. We find
\begin{equation}
    S_L^{(N)}(N\bar{q})= N\left[\log 2+ \bar{q}\log \frac{1-2\bar{q}}{1+2\bar{q}}-\frac{1}{2}\log\left(1-4 \bar{q}^2\right)\right]+\mathcal{O}(\log N)\,.
\end{equation}
When evaluating this for $\bar{q}=\frac{1}{2}\frac{1-\lambda}{1+\lambda}$, we find
\begin{equation}
    S_{L}^{(N)}(N\bar{q})\big|_{\bar{q}=\frac{1}{2}\frac{1-\lambda}{1+\lambda}}=-N\left(-\frac{1}{2}\frac{1-\lambda}{1+\lambda}\log{\lambda}+\log{\frac{\sqrt{\lambda}}{1+\lambda}}\right)+\mathcal{O}(\log N)\,.
\end{equation}
Comparing with the full entanglement entropy \eqref{eq:full_entanglement_entropy_finite_N}, we find that 
\begin{equation}
    S_L^{(N)}= S_{L}^{(N)}(N\bar{q})\big|_{\bar{q}=\frac{1}{2}\frac{1-\lambda}{1+\lambda}}+\mathcal{O}(\log N)\,.
\end{equation}
Thus, up to logarithmic corrections in $N$, the entanglement entropy receives contributions only by the charge sector with $\bar{q}=\frac{1}{2}\frac{1-\lambda}{1+\lambda}$, consistent with the peak of the probability distribution discussed above.

\section{Symmetry resolution of hyperfinite algebras}
\label{sec:hyeprfiniteresolved}

\subsection{General strategy}


A central result of this paper is the construction of symmetry-resolved hyperfinite algebras by combining subalgebras associated to different values of the subregion charge. As we explain below, 
our approach relies on the use of \textit{direct integrals}. Indeed, the direct sum of factors and its continuum limit given by the direct integral is, by construction, an algebra endowed with sectors (see section 3 of \cite{Sorce:2023fdx}).
As we concluded in \secref{subsec:rescaling}, introducing a charge in the setup described in \secref{subsec:rescaling} leads to a subregion charge $s_L$ in \eqref{eq:rescaled_charge_typeIII_lambda_limit}  proportional to the identity and, therefore, to the local observable algebras trivially coinciding with the local field algebras, i.e.~$\mathcal{F}(L)=\mathcal{A}(L)$. 
In what follows, we use these algebras for different values of the subregion charge as building blocks for the symmetry resolution of hyperfinite algebras. To this end, we label the local algebras with the unique eigenvalue of the subregion charge operator $s_L$.

We notice that, due to the rescaling by $N$, the values of subregion charge $s_L$ continuously vary in the set $[-1/2,1/2]$. Thus, in the most general scenario, the direct sum over the fixed-subregion-charge factors has to run over a continuous set. In mathematical terms, this means that we have to resort to direct integrals. In fact, in QFT, there are instances of operators with continuous spectra. Decomposing these algebras over the eigenvalues of such operators, generalizing what we usually do for conserved charges, requires the use of direct integrals. In order to keep our new proposal as general as possible, we thus utilize direct integrals.

Below we review direct integrals of von Neumann algebras and use these tools to construct algebras resolved into hyperfinite factors associated with eigenvalues of the subregion charge.
The subregion charge $s_L$ has a unique (continuous) eigenvalue $\bar{q}$ determined by the vector $\boldsymbol{\lambda}$, as defined in \eqref{eq:rescaled_charge_typeIII_lambda_limit} and \eqref{eq:rescaled_charge_type_III1}. Thus, we introduce the dependence on the subregion charge eigenvalue of the local algebras as
 $\mathcal{A}_{\bar{q}}(L)$. In the setup considered in this manuscript, $\boldsymbol{\lambda}$ also specifies the type of the factor. As we can tune $\boldsymbol{\lambda}$ to access various types of hyperfinite algebras, the direct integral over $\mathcal{A}_{\bar{q}}(L)$ provides a resolution of a (larger) local algebra into hyperfinite algebras. This is one of the main goals of this manuscript. In \secref{sec:modflow_Resolu}, we use this explicit construction to study the symmetry resolution of the modular operator and modular flow in hyperfinite algebras, which is the second main objective of this paper. While in the present work, we focus on the case of symmetry groups generated by a single charge, the construction can be readily extended to non-Abelian symmetries to generalize the construction developed in \cite{Bianchi:2024aim} for type I algebras. These aspects are discussed further in \appref{app:non_abelian_and_discrete_resolution}.

\subsection{Direct integrals of von Neumann algebras}\label{sec:direct_integrals_review}

This review of direct integrals closely follows Volume I of \cite{Takesaki1979} and the reader familiar with this subject may skip this section. The physical intuition behind the concept of direct integrals amounts to a continuum limit of direct sums. 

Let us start with the definition of a direct integral of a family of Hilbert spaces. Let $(X,\Sigma, \mu)$ be a measure space, where $X$ is a set, $\Sigma$ is a $\sigma$-algebra on $X$ and $\mu$ is a measure. Let $\mathcal{H}(\alpha)$ be a family of Hilbert spaces indexed by $\alpha\in X$. Before defining the direct integral, we need to define the Cartesian product $\prod_{\alpha\in X} \mathcal{H}(\alpha)$ of these Hilbert spaces. This mirrors the definition of the direct sum which is defined as a certain subset of the tensor product.
In the case of continuous $X$, we define $\prod_{\alpha\in X} \mathcal{H}(\alpha)$ as the set of maps $\ket{\mathfrak{f}}$ such that
\begin{equation}\label{eq:cartesian_product_definition}
    \ket{\mathfrak{f}}:X\rightarrow\bigcup_{\alpha\in X} \mathcal{H}(\alpha), \qquad \alpha\rightarrow \ket{\mathfrak{f}}(\alpha)=\ket{f(\alpha)}\in \mathcal{H}(\alpha).
\end{equation}
Intuitively, we can think of $\mathcal{H}(\alpha)$ as fibers along $X$ and $\ket{\mathfrak{f}}$ as sections along the fibers. We now consider a subset $\mathfrak{G}\subset\prod_{\alpha\in X} \mathcal{H}(\alpha)$ of the Cartesian product. We call $(\{\mathcal{H}(\alpha)\},\mathfrak{G})$ a measurable field of Hilbert spaces if the following conditions are satisfied.
    \begin{itemize}
        \item[(i)]
        For any $\ket{\mathfrak{g}}\in \mathfrak{G}$ the function $\alpha\in X\rightarrow \braket{g(\alpha)}^{\frac{1}{2}}\in\mathbb{R}$ is measurable;
        \item[(ii)]  For any $\ket{\mathfrak{p}}\in \prod_{\alpha\in X} \mathcal{H}(\alpha)$, if the function $\alpha\rightarrow\braket{g(\alpha)}{p(\alpha)}$ is measurable for every $\ket{\mathfrak{g}}\in \mathfrak{G}$, then $\ket{\mathfrak{p}}\in \mathfrak{G}$;
        \item[(iii)]
There exists a countable subset $\{\ket{\mathfrak{g}_1},\ket{\mathfrak{g}_2},\dots\}$ of $\mathfrak{G}$ such that for all $\alpha\in X$ the Hilbert space $\mathcal{H}(\alpha)$ is the closed span of $\{\ket{g_n(\alpha)}\}$.
    \end{itemize}
In this case, elements of $\mathfrak{G}$ are called measurable vector fields. The direct integral $\mathfrak{H}$ of the $\mathcal{H}(\alpha)$ is defined as the set of measurable vector fields such that
\begin{equation}
        \braket{\mathfrak{g}}\equiv\int_X \braket{g(\alpha)}\mathrm{d}\mu(\alpha)<\infty\,.
\end{equation}
Here, the bracket inside the integral denotes the inner product on $\mathcal{H}(\alpha)$. More precisely, we identify vector fields that differ on subsets of $X$ with measure $0$ and consider equivalence classes under this relation.
We denote the direct integral as
    \begin{equation}\label{eq:def_direct_int_hilbert_space}
        \mathfrak{H}=\int_X^\oplus \mathcal{H}(\alpha) \mathrm{d}\mu(\alpha)\,.
    \end{equation}
Crucially, $\mathfrak{H}$ is again a Hilbert space with inner product for $\ket{\mathfrak{g}},\ket{\mathfrak{f}}\in\mathfrak{H}$ given by
    \begin{equation}\label{eq:direct_int_hilbert_space_inner_product}
        \braket{\mathfrak{g}}{\mathfrak{f}}\equiv\int_X \braket{g(\alpha)}{f(\alpha)}\mathrm{d}\mu(\alpha)\,.
\end{equation}

To gain an intuition for these definitions, consider the following example. Let $\mathcal{H}(\alpha)=\mathrm{span}(\ket{\uparrow}_\alpha,\ket{\downarrow}_\alpha)$, where $\alpha\in[0,1]$. The Cartesian product $\prod_\alpha \mathcal{H}(\alpha)$ can then be parametrized by maps
\begin{equation}
    \ket{f(\alpha)}=c_1(\alpha)\ket{\uparrow}_\alpha+c_2(\alpha)\ket{\downarrow}_\alpha,\quad c_1, c_2: [0,1]\rightarrow\mathbb{C}.
\end{equation}
The maps $c_1,c_2$ need not necessarily be measurable, i.e.~they could be indicator functions of non-measurable subsets of $[0,1]$. To define the set of measurable vector fields, we only consider measurable functions,
\begin{equation}\label{eq:measurable_vector_field_2_spins_example}
    \mathfrak{G}=\left\{\ket{\mathfrak{g}}\in\prod_\alpha \mathcal{H}(\alpha)\;|\;\ket{g(\alpha)}=m_1(\alpha)\ket{\uparrow}_\alpha+m_2(\alpha)\ket{\downarrow}_\alpha,\;m_1,m_2\text{ measurable}\right\}.
\end{equation}
This subset $\mathfrak{G}$ of the Cartesian product satisfies axioms (i)-(iii). (i) is satisfied, since the expression $\braket{g(\alpha)}{g(\alpha)}^{\frac{1}{2}}=\sqrt{m_1^*m_1+m_2^*m_2}$ is measurable, which follows from the fact that products, sums and compositions of measurable functions are measurable. (iii) is satisfied, since the set $\{\ket{\mathfrak{g}_1},\ket{\mathfrak{g}_2}\}$ with $\ket{g_1(\alpha)}=\ket{\uparrow}_\alpha$ and $\ket{g_2(\alpha)}=\ket{\downarrow}_\alpha$ pointwisely spans $\mathcal{H}(\alpha)$. To show (ii) consider $\ket{\mathfrak{p}}\in\prod_\alpha\mathcal{H}(\alpha)$ with $\ket{p(\alpha)}=p_1(\alpha)\ket{\uparrow}_\alpha+p_2(\alpha)\ket{\downarrow}_\alpha$. We now have to show, that, given measurability of the function $\braket{g(\alpha)}{p(\alpha)}$ for any $\ket{\mathfrak{g}}\in\mathfrak{G}$, then also $p_1, p_2$ are measurable, i.e.~$\ket{\mathfrak{p}}\in\mathfrak{G}$. To show this, consider $m_1=1$, $m_2=1$, such that $f_1=p_1+p_2$ is measurable. Consider further $m_1=1$ and $m_2=-1$, such that $f_2=p_1-p_2$ is measurable. Thus, we can always write $p_1$ and $p_2$ as linear combinations of measurable functions, and thus $\ket{\mathfrak{p}}\in\mathfrak{G}$. The direct integral now consists of vector fields
\begin{equation}
\begin{split}
    \int_{[0,1]}^\oplus \mathrm{d}\mu(\alpha) \mathcal{H}(\alpha) = &\{\ket{\mathfrak{g}}\,\vert\, \ket{g(\alpha)}=m_1(\alpha)\ket{\uparrow}_\alpha+m_2(\alpha)\ket{\downarrow}_\alpha\}\,, 
\end{split}
\end{equation}
where $m_1$ and $m_2$ are measurable and square integrable. More precisely, it consists of equivalence classes of these vector fields obtained by identification of functions that are the same almost everywhere, i.e.~differ at most on sets of measure $0$. The inner product of $\ket{\mathfrak{g}}, \ket{\mathfrak{f}}$ with $\ket{g(\alpha)}=m_1(\alpha)\ket{\uparrow}_\alpha+m_2(\alpha)\ket{\downarrow}_\alpha$ and $\ket{f(\alpha)}=m'_1(\alpha)\ket{\uparrow}_\alpha+m'_2(\alpha)\ket{\downarrow}_\alpha$ is then given by
\begin{equation}
    \braket{\mathfrak{g}}{\mathfrak{f}}=\int_{[0,1]} \mathrm{d}\mu(\alpha) (m_1^*m_1'+m_2^* m_2').
\end{equation}
In the case of a discrete index set $X$, we choose $\mu$ to be the counting measure. In this case, these definitions straightforwardly lead to a direct sum of Hilbert spaces. Thus, we may regard this construction of direct integrals as an extension of direct sums for continuous index sets.

Having defined direct integrals of Hilbert spaces, we now turn to the respective algebras. Let us denote by $\mathcal{A}_\alpha$ a family of von Neumann algebras on $\mathcal{H}(\alpha)$. We define the Cartesian product analogously to \eqref{eq:cartesian_product_definition} and denote its elements by $\mathfrak{a}: \alpha\in X\rightarrow a(\alpha)\in\mathcal{A}_\alpha$. Such an operator field is called measurable, if, for all $\ket{\mathfrak{s}}\in\mathfrak{H}$, the vector field $\mathfrak{a}\ket{\mathfrak{s}}:\alpha\rightarrow a(\alpha)\ket{s(\alpha)}$ is measurable. If $||a(\alpha)||<\infty$ almost everywhere, then $\mathfrak{a}\ket{\mathfrak{s}}\in\mathfrak{H}$ and we denote
\begin{equation}\label{eq:decomposable_operator}
    \mathfrak{a}=\int_X^\oplus a(\alpha) \mathrm{d}\mu(\alpha)\,.
\end{equation}
Operators of this form are called decomposable. If there exists a countable set $\{\mathfrak{a}_n\}$ of measurable operators that pointwisely generate $\mathcal{A}_\alpha$ for almost all $\alpha\in X$, we call the family $\mathcal{A}_\alpha$ measurable. In this case, we call the set of decomposable operators \eqref{eq:decomposable_operator} the direct integral of $\mathcal{A}_\alpha$ and denote it by
\begin{equation}\label{eq:def_direct_int_algebra}
    \mathfrak{A}=\int_X^\oplus \mathcal{A}(\alpha) \mathrm{d}\mu(\alpha)\,.
\end{equation}
$\mathfrak{A}$ is a von Neumann algebra, and its commutant is given by
\begin{equation}
    \mathfrak{A}'=\int_X^\oplus \mathcal{A}'_\alpha\mathrm{d}\mu(\alpha)\,,
\end{equation}
where $\mathcal{A}'_\alpha$ is the commutant of $\mathcal{A}_\alpha$.
The adjoint operator with respect to the inner product \eqref{eq:direct_int_hilbert_space_inner_product} is defined as
\begin{equation}\label{eq:defintion_adjoint}
    \mathfrak{a}^\dagger=\int^\oplus_X\mathrm{d}\mu(\alpha)\,a(\alpha)^\dagger\,,   
\end{equation}
where $a(\alpha)^\dagger$ is the adjoint operator with respect to the inner product in $\mathcal{H}(\alpha)$.
Importantly, the direct integral $\mathfrak{A}$ is \textbf{not a factor}. Its centre contains operators of the form \cite{chow1970spectral}
\begin{equation}\label{eq:central_elements_direct_integral}
    \mathfrak{c}=\int_X^\oplus c(\alpha)\mathrm{id}(\alpha)\mathrm{d}\mu (\alpha)\,,
\end{equation}
where $c(\alpha)$ is a essentially bounded scalar function, and $\mathrm{id}(\alpha)$ is the identity in $\mathcal{H}(\alpha)$. Note that operators of this form are not proportional to the identity operator on $\mathfrak{H}$. In the finite dimensional case, they correspond to diagonal matrices with different eigenvalues. The fact that operators of the above form span the centre is of great importance for the following discussion. It may be understood in analogy to the direct sum case, where the centre is given by linear combinations of the projection operators. The operators $\mathrm{id}(\alpha)$, in a sense which will be made precise in \secref{sec:fixed_charge_algebras}, may be regarded as analogous to the projection operators, when the direct sum is again replaced by the direct integral.

\subsection{Direct integrals and symmetry resolution}\label{sec:fixed_charge_algebras}

We now make use of these properties of the direct integral, in particular \eqref{eq:central_elements_direct_integral}, to construct hyperfinite subregion algebras with a central element which we interpret as a subregion charge. 
For simplicity, we restrict our analysis to the case where all the $\mathcal{A}_\alpha$ are of the same type, which also defines the type of the direct integral algebra.
For the labels, we now use the eigenvalues of the operators $s_L$, i.e.~$\bar{q}\in X= [-\frac{1}{2},\frac{1}{2}]$, as discussed in \secref{subsec:rescaling}.
Moreover, we reinstate the explicit dependence of the algebras on the subregion of the algebras as in \eqref{eq:subregion_algebra_itpfi}.
Accordingly, we denote the algebras by $\mathcal{A}_{\bar{q}}(L)$.

Our starting point is the direct integral of the Hilbert spaces \eqref{eq:AW_Hilbert_space_definition}. Recalling the relations \eqref{eq:rescaled_charge_typeIII_lambda_limit} and \eqref{eq:rescaled_charge_type_III1}, we see that the properties of the cyclic separating state \eqref{eq:cyclicseparating AW} determine the eigenvalue of $s_L$ associated with the particular Hilbert space, labelled again by $\mathcal{H}(\bar{q})$.\footnote{This notation for the Hilbert space should not be confused with the one introduced in \eqref{eq:SSSdecomposition} to denote the fixed superselection sector Hilbert space.} At first, we have to check whether the axioms (i)-(iii) in \secref{sec:direct_integrals_review} are satisfied. Similarly to the example presented in \secref{sec:direct_integrals_review}, we define $\mathfrak{G}$ as in \eqref{eq:measurable_vector_field_2_spins_example}, except that now $\ket{g(\alpha)}\rightarrow\ket{\Psi({\bar{q})}}$ is a generic vector of $\mathcal{H}(\bar{q})$.
In particular, due to the infinite dimensionality of $\mathcal{H}(\bar{q})$, $\ket{\Psi({\bar{q})}}$ can be written as combination of infinitely many vectors with coefficients being measurable functions of $\bar{q}$. Using that infinite sums of measurable functions are still measurable, we conclude that
 that (i) and (ii) are satisfied. Note that we can explicitly construct a set of vectors satisfying (iii) by considering
\begin{equation}\label{eq:fundamental_vectors}
    v_1\otimes\dots \otimes v_k\otimes K_{2,\lambda_{k+1}}\otimes K_{2,\lambda_{k+2}}\otimes\dots \in \mathcal{H}(\bar{q})\,,
\end{equation}
where
\begin{equation}
    v_i\in \left\{\begin{pmatrix}
        1 & 0\\
        0&0
    \end{pmatrix}, \begin{pmatrix}
        0&1\\
        0&0
    \end{pmatrix}, \begin{pmatrix}
        0&0\\
        1&0
    \end{pmatrix},\begin{pmatrix}
        0&0\\
        0&1
    \end{pmatrix}\right\}\,.
\end{equation}
We denote the direct integral as
\begin{equation}
    \mathfrak{H}=\int_{[-\frac{1}{2},\frac{1}{2}]}^\oplus \mathrm{d}\mu(\bar{q}) \;\mathcal{H}(\bar{q})\,.
\end{equation}

We now construct the direct integral algebra and study its properties. In particular, we define a coarse- and fine-grained symmetry resolution and corresponding projection operators.
We consider the direct integral algebra
\begin{equation}
    \mathfrak{A}(L)\equiv \int_{[-\frac{1}{2},\frac{1}{2}]}^\oplus \mathrm{d}\mu(\bar{q}) \;\mathcal{A}_{\bar{q}}(L)\,.
\label{eq:SRalgebra_directint}
\end{equation}
We emphasize here that in general the algebras $\mathcal{A}_{\bar{q}}(L)$ and $\mathfrak{A}(L)$ act on vastly different Hilbert spaces, $\mathcal{A}_{\bar{q}}(L)$ acting on $\mathcal{H}(\bar{q})$, while $\mathfrak{A}(L)$ acts on $\mathfrak{H}$. Intuitively, this can be understood by noticing that the action of $\mathcal{A}_{\bar{q}}(L)$ would only change the vector in $\mathfrak{H}$ at a single point, which, in general, is a measure zero subset of $X$. This is different from the type I/direct sum case, where, by construction, the constituent symmetry-resolved algebras have an action on the Hilbert space of the full invariant algebra acts. 
Despite this fact, the direct integral construction is as a generalization of the type I case, since, as discussed in \secref{sec:direct_integrals_review}, it reduces to a direct sum when $X$ has a discrete topology, which is the case discussed in \secref{sec:type_I_resolution}.

We continue by examining the structure of the algebra $\mathfrak{A}(L)$ and in particular its centre. By construction, the operator
\begin{equation}
    \mathfrak{s}_L=\int_{[-\frac{1}{2},\frac{1}{2}]}^\oplus \mathrm{d}\mu(\bar{q})\, s_L\,
\end{equation}
is an element of the centre of $\mathfrak{A}(L)$. From now on, we suppress the integration domain, which we from now on assume to be $X=[-1/2,1/2]$. In fact, we should think of the centre in a way similar to the type I case, where the centre consists of linear superpositions of projection operators. In the direct integral case, projectors may be defined as
\begin{equation}\label{eq:projectors_fine_grained}
    \Tilde{\Pi}(\bar{q}): \mathfrak{H}\to \mathcal{H}(\bar{q})\,,\qquad \ket{\mathfrak{g}}=\int^\oplus\mathrm{d}\mu(\bar{q})\,\ket{g(\bar{q})}\mapsto \ket{g(\bar{q})}\,.
\end{equation}
Note that these projection operators are not operators on $\mathfrak{H}$ and thus not elements of the algebra $\mathfrak{A}(L)$. This is due to the fact that these operators project only to one respective charge sector, which is a measure zero subset. Therefore these operators are zero almost everywhere. Thus, from the point of view of the direct integral Hilbert space, they may be identified with the zero operator. 
We also define a projector from the dual space, which we denote by
\begin{equation}\label{eq:projectors_fine_grained_adjoint}
    \Tilde{\Pi}^\dagger(\bar{q}):\mathfrak{H}^*\to \mathcal{H}(\bar{q})^*\,,\qquad \bra{\mathfrak{g}}=\int^\oplus\mathrm{d}\mu(\bar{q})\,\bra{g(\bar{q})}\mapsto \bra{g(\bar{q})}\,.
\end{equation}

There is a different notion of projectors associated with a subset $E$ of non-zero measure of the interval $X=[-1/2,1/2]$,
\begin{equation}
    \Pi(E): \mathfrak{H}\to\mathfrak{H}, \qquad  \ket{\mathfrak{g}}=\int^\oplus\mathrm{d}\mu(\bar{q})\,\ket{g(\bar{q})}\mapsto \Pi(E)\ket{\mathfrak{g}}\equiv\int^\oplus \mathrm{d}\mu(\bar{q})\, \mathcal{I}(E) \ket{g(\bar{q})}\,,
\end{equation}
where $\mathcal{I}(E)$ is the indicator function of $E$. The operators $\Pi(E)$ are proper elements of the direct integral algebra $\mathfrak{A}(L)$ and can be decomposed as
\begin{equation}\label{eq:projectors_coarse_grained}
    \Pi(E)=\int^\oplus\mathrm{d}\mu(\bar{q}) \,\mathcal{I}(E) \,\mathrm{id}(\bar{q})\,.
\end{equation}
It may be shown straightforwardly that the set of these projectors defines a projection-valued measure. In particular, because $\mathcal{I}(E)^2=\mathcal{I}(E)$, $\Pi(E)$ is idempotent. Moreover, $\Pi(E)$ commutes with all operators in $\mathfrak{A}(L)$, since it is of the form \eqref{eq:central_elements_direct_integral}.

Now, we may approximate any central elements of the form \eqref{eq:central_elements_direct_integral} by a sequence of successively smaller, disjoint, intervals $E_k$, and a sequence $c_k$, 
\begin{equation}
    \mathfrak{c}=\lim_{n\to\infty} \sum_{k=1}^n c_k \Pi(E_k)\,,
\end{equation}
in a similar way to how we approximate any integrable function,
\begin{equation}
    c(\alpha)=\lim_{n\to\infty} \sum_{k=1}^n c_k \mathcal{I}(E_k)\,.
\end{equation}
This realizes symmetry resolution as the centre of the algebra is the span of the projectors $\Pi(E)$. This is one of the main results of this paper. It implies that the full spectrum of charges can be accessed by a sequence of projectors in the the local algebra.

In summary, we find that there are two possible ways to define projection operators on the Hilbert space $\mathfrak{H}$. The projectors \eqref{eq:projectors_coarse_grained} may be understood as a coarse graining of the subregion charge, whereas the projectors in \eqref{eq:projectors_fine_grained} in this picture correspond to a fine graining. In defining the coarse-grained projectors, we sacrifice precision in the knowledge of the subregion charge. However, the projectors we obtain this way are mathematically better behaved, as they are elements of $\mathfrak{A}(L)$, whereas the fine-grained projectors are not. 

A similar notion of coarse-graining also exists for the algebras. Instead of considering the algebras $\mathcal{A}_{\bar{q}}(L)$, the fine-grained algebras, we instead consider the coarse-grained algebras associated with $E$ as
\begin{equation}
    \mathcal{A}_E(L)=\int^\oplus \mathrm{d}\mu(\bar{q}) \mathcal{I}(E) \mathcal{A}_{\bar{q}}(L)\,.
\end{equation}
In contrast to the algebras $\mathcal{A}_{\bar{q}}(L)$, the coarse-grained algebra acts on the same Hilbert space $\mathfrak{H}$ as the algebra $\mathfrak{A}(L)$. If we assume a collection of disjoint intervals $E_i$ such that their union is $[-1/2,1/2]$, we may define a resolution
\begin{equation}
\label{eq:coarse-grained resolution}
    \mathfrak{A}(L)=\bigoplus_i \mathcal{A}_{E_i}(L)\,.
\end{equation}
We note that this construction has a different structure than the fine-grained resolution, as the algebras $\mathcal{A}_{E_i}(L)$ themselves have a centre.

The coarse-grained resolution exhibits intriguing similarities with the regularization procedure for the charge sectors applied in the standard literature on symmetry-resolved entanglement in CFT, as reviewed in \secref{subsec:rescaling}.
Indeed, as we see in \eqref{eq:sr_rescaled_charge}, the set of allowed values of rescaled charge $\tilde{q}$ becomes more and more dense as the UV cutoff $\varepsilon$ approaches zero. However, as $\varepsilon$ must be kept small but finite, the resulting set of values of $\tilde{q}$ has a spacing which depends on the choice of $\varepsilon$. In other words, different choices of $\varepsilon$ correspond to different coarse-grainings of the charge values. In this sense, it is similar to considering different instances of intervals $E_k$, leading to distinct coarse-grained resolutions \eqref{eq:coarse-grained resolution}.
For this reason, the coarse-grained resolution could be potentially used to frame results established in the entanglement literature in a mathematically solid manner.

We conclude by observing that, in principle, we may calculate the entropies associated with a cyclic separating state in the coarse- or fine-grained algebras to obtain a symmetry resolution of entanglement. However, we would still obtain divergent quantities as both the coarse and fine-grained algebras are hyperfinite.

\section{Symmetry resolution of hyperfinite modular theory}
\label{sec:modflow_Resolu}

We now use the methods developed in \secref{sec:hyeprfiniteresolved} to examine the symmetry resolution of the modular operator, modular flow, and modular correlation functions defined for hyperfinite algebras. We show that the symmetry-resolved modular correlation functions satisfy the KMS condition, which is one of the main results of this work.

\subsection{Resolved modular operator and modular flows}

We start by defining the modular operator in the direct integral algebra $\mathfrak{A}(L)$. In order to define a Tomita operator similar to \eqref{eq:Modular relation general}, we first need to find a cyclic separating vector. To this end, we define the state vector
\begin{equation}\label{eq:cyclic_separating_vector_direct_integral}
    \ket{\mathfrak{P}}=\int^\oplus\mathrm{d}\mu(\bar{q}) \,\sqrt{p(\bar{q})}\,\ket{\Psi_{\boldsymbol{\lambda}}}\,,
\end{equation}
where $\ket{\Psi_{\boldsymbol{\lambda}}}\in \mathcal{H}_{\boldsymbol{\lambda}}$ is the cylic separating vector defined in \eqref{eq:cyclicseparating AW} and $p(\bar{q})$ is a probability distribution, which we assume to be non-zero almost everywhere. Thus,
\begin{equation}
    \ket{\Psi_{\boldsymbol{\lambda}}}=\frac{1}{\sqrt{p(\bar{q})}}\tilde{\Pi}(\bar{q})\ket{\mathfrak{P}}\,,
\end{equation}
where $\tilde{\Pi}$ is defined in \eqref{eq:projectors_fine_grained}.
To verify that $\ket{\mathfrak{P}}$ is indeed cyclic and separating for $\mathfrak{A}(L)$, we note that by definition of the direct integral algebra \eqref{eq:def_direct_int_algebra} and the direct integral Hilbert space \eqref{eq:def_direct_int_hilbert_space},
\begin{equation}
    \overline{\mathfrak{A}(L)\ket{\mathfrak{P}}}=\mathfrak{H}\,.
\end{equation}
Thus, $\ket{\mathfrak{P}}$ is cyclic. We show that it is separating by contradiction. To this end, assume that there exists $\mathfrak{a}\in \mathfrak{A}(L)$, such that
\begin{equation}
    \mathfrak{a}\ket{\mathfrak{P}}=\int^\oplus\mathrm{d}\mu(\bar{q}) \, \sqrt{p(\bar{q})} a(\bar{q}) \ket{\Psi_{\boldsymbol{\lambda}}} \overset{!}{=}0\,.
\end{equation}
In order for this equation to be true, $a(\bar{q}) \ket{\Psi_{\boldsymbol{\lambda}}}=0$ almost everywhere.
We also know that the states $\ket{\Psi_{\boldsymbol{\lambda}}}$ are separating in the fixed charge Hilbert spaces, i.e.~$a(\bar{q})\ket{\Psi_{\boldsymbol{\lambda}}}=0\Leftrightarrow a(\bar{q})=0$. Putting together those two facts, we have that $a(\bar{q})=0$ almost everywhere. Since we identify operators that are equal almost everywhere, we find that
\begin{eqnarray}
    \mathfrak{a}\ket{\mathfrak{P}}=0\,\quad \Leftrightarrow \quad \mathfrak{a}=0\,.
\end{eqnarray}
We conclude that indeed $\ket{\mathfrak{P}}$ is a cyclic separating vector. Under these conditions, we can define the Tomita operator associated to the vector $\ket{\mathfrak{P}}$ and the algebra $\mathfrak{A}(L)$ as
\begin{equation}
    \mathfrak{S}_{\mathfrak{P}}=\int^\oplus\mathrm{d}\mu(\bar{q}) S_{\Psi_{\boldsymbol{\lambda}}}(\bar{q})\,,
\end{equation}
where $S_{\Psi_{\boldsymbol{\lambda}}}(\bar{q})$ is the Tomita operator associated with $\ket{\Psi_{\boldsymbol{\lambda}}}$ and the algebra $\mathcal{A}_{\bar{q}}(L)$. This operator satisfies \eqref{eq:Modular relation general} as can be checked by using the definition \eqref{eq:defintion_adjoint},
\begin{equation}
    \mathfrak{S}_\mathfrak{P}\mathfrak{a}\ket{\mathfrak{P}}= \int^\oplus\mathrm{d}\mu(\bar{q}) \,\sqrt{p(\bar{q})}S_{\Psi_{\boldsymbol{\lambda}}}a(\bar{q}) \ket{\Psi_{\boldsymbol{\lambda}}}=\int^\oplus\mathrm{d}\mu(\bar{q}) \,\sqrt{p(\bar{q})}a(\bar{q})^\dagger \ket{\Psi_{\boldsymbol{\lambda}}}=\mathfrak{a}^\dagger\ket{\mathfrak{P}}\,.
\end{equation}
Similarly to \eqref{eq:polardecompositiongeneral}, we decompose the Tomita operator as
\begin{equation}
    \mathfrak{S}_\mathfrak{P}=\mathfrak{J}_\mathfrak{P}\Delta_\mathfrak{P}^{\frac{1}{2}}\,,
\end{equation}
where
\begin{equation}\label{eq:mod_op_direct_integral}
    \Delta_\mathfrak{P}=\mathfrak{S}^\dagger_\mathfrak{P}\mathfrak{S}_\mathfrak{P}=\int^\oplus\mathrm{d}\mu(\bar{q}) \,S_{\Psi_{\boldsymbol{\lambda}}}(\bar{q})^\dagger S_{\Psi_{\boldsymbol{\lambda}}}(\bar{q})=\int^\oplus\mathrm{d}\mu(\bar{q}) \,\Delta_{\Psi_{\boldsymbol{\lambda}}}\,,
\end{equation}
where $\Delta_{\Psi_{\boldsymbol{\lambda}}}$ is the modular operator associated with $\ket{\Psi_{\boldsymbol{\lambda}}}$ and the algebra $\mathcal{A}_{\bar{q}}(L)$. Thus, we write the modular operator $\Delta_\mathfrak{P}$ associated to the algebra $\mathfrak{A}(L)$ as a direct integral of modular operators associated with the algebras $\mathcal{A}_{\bar{q}}(L)$. Thus, in \eqref{eq:mod_op_direct_integral} we have found a generalization of \eqref{eq:SRdec_modular_operator} by writing down a decomposition of the modular operator into parts associated with fixed subregion charge.

Note that because neither $S_{\Psi_{\boldsymbol{\lambda}}}$ nor $\Delta_{\Psi_{\boldsymbol{\lambda}}}$ are elements of $\mathcal{A}_{\bar{q}}(L)$ in general, $\mathfrak{S}_\mathfrak{P}$ and $\Delta_\mathfrak{P}$ are also not elements of $\mathfrak{A}(L)$, as expected for a modular operator. 
Given that the modular operators $\Delta_{\Psi_{\boldsymbol{\lambda}}}$ satisfy the Tomita-Takesaki theorem \eqref{eq:TTtheorem} for the algebra $\mathcal{A}_{\bar{q}}(L)$, the modular operator $\Delta_\mathfrak{P}$ satisfies the Tomita-Takesaki theorem for the algebra $\mathfrak{A}(L)$.

Analogously to \eqref{eq:cyclic_separating_vector_direct_integral}, \eqref{eq:mod_op_direct_integral}, we now define cyclic separating vector, modular operator and probabilities associated to the coarse grained algebra as discussed in \secref{sec:fixed_charge_algebras},
\begin{align}
    \ket{\Psi_E}&\equiv\Pi(E)\ket{\mathfrak{P}}= \int^\oplus\mathrm{d}\mu(\bar{q})\,\mathcal{I}(E)\sqrt{p(\bar{q})}\ket{\Psi_{\boldsymbol{\lambda}}}\,,\label{eq:coarse_grained_cyclic_sep_vector}\\
     \Delta_{\Psi_E}&\equiv\Pi(E)\Delta_\mathfrak{P}= \int^\oplus\mathrm{d}\mu(\bar{q})\,\mathcal{I}(E)\Delta_{\Psi_{\boldsymbol{\lambda}}}\,,\label{eq:coarse_grained_mod_op}\\
    p(E)&\equiv \braket{\Psi_E}{\Psi_E}=\int\mathrm{d}\mu(\bar{q})\,\mathcal{I}(E) p(\bar{q})\,.
\end{align}
Here, $p(E)$ is the probability of measuring subregion charge $\bar{q}\in E$. Upon choosing a disjoint collection $E_i$ whose union is $X$, we recover the normalization $\sum_i p(E_i)=1$.
By following similar steps as above,
it is straightforward to show that $\ket{\Psi_E}$ is a cyclic separating vector associated to the algebra $\mathcal{A}_E(L)$.

Paralleling the argument in \secref{sec:type_I_resolution}, the symmetry-resolutions of the modular operator \eqref{eq:mod_op_direct_integral} and \eqref{eq:coarse_grained_mod_op} imply the resolution of the modular flow \eqref{eq:TTtheorem}. This is a further extension of the results in \cite{Di_Giulio_2023} (where only the type I case was considered) provided by our investigation.

\subsection{
Symmetry-resolved modular correlation functions and KMS condition}

Having defined the relevant operators, we now define a symmetry-resolution of modular correlation functions of operators in the algebra $\mathfrak{A}(L)$. In analogy to \eqref{eq:modcorrfunc_AQFT}, we define the modular correlation function for $\mathfrak{a},\mathfrak{b}\in\mathfrak{A}(L)$ as
\begin{equation}\label{eq:full_mod_correlation_function}
     G_{\textrm{\tiny mod}} (\mathfrak{a},\mathfrak{b};t)\equiv\bra{\mathfrak{P}}\mathfrak{b}\Delta_\mathfrak{P}^{i t}\mathfrak{a}\Delta_\mathfrak{P}^{-it}\ket{\mathfrak{P}}\,.
\end{equation}
Using the decomposition \eqref{eq:mod_op_direct_integral}, as well as the decomposition of the cyclic separating vector \eqref{eq:cyclic_separating_vector_direct_integral}, we find
\begin{equation}\label{eq:fine_grained_modular_correlation_function}
\begin{split}
    G_{\textrm{\tiny mod}} (\mathfrak{a},\mathfrak{b};t) &=\int\mathrm{d}\mu(\bar{q}) \,p(\bar{q})\bra{\Psi_{\boldsymbol{\lambda}}} b(\bar{q}) \Delta_{\Psi_{\boldsymbol{\lambda}}}^{it}a(\bar{q})\Delta_{\Psi_{\boldsymbol{\lambda}}}^{-it}\ket{\Psi_{\boldsymbol{\lambda}}} \\
    &\equiv\int\mathrm{d}\mu(\bar{q})\,p(\bar{q})\,G_{\textrm{\tiny mod}} (a(\bar{q}),b(\bar{q});t,\bar{q})\,.
\end{split}
\end{equation}
Here, in analogy to \eqref{eq:modular_correlation_function_decomposition}, we defined the symmetry-resolved modular correlation functions,
\begin{equation}\label{eq:fine_grained_modular_correlation_function_def}
    G_{\textrm{\tiny mod}} (a(\bar{q}),b(\bar{q});t,\bar{q})\equiv \bra{\Psi_{\boldsymbol{\lambda}}} b(\bar{q}) \Delta_{\Psi_{\boldsymbol{\lambda}}}^{it}a(\bar{q})\Delta_{\Psi_{\boldsymbol{\lambda}}}^{-it}\ket{\Psi_{\boldsymbol{\lambda}}}\,.
\end{equation}
We may also define a coarse-grained version of the modular correlation function associated to the coarse-grained algebras $\mathcal{A}(E)$, c.f. the discussion in \secref{sec:fixed_charge_algebras},
\begin{equation}\label{eq:coarse_grained_modular_correlation_function}
    G_{\textrm{\tiny mod}} (a(E),b(E);t,E)\equiv \bra{\Psi_E}b(E)\Delta_{\Psi_E}^{it}a(E)\Delta_{\Psi_E}^{-it}\ket{\Psi_E}\,.
\end{equation}
From \eqref{eq:coarse_grained_cyclic_sep_vector} and \eqref{eq:coarse_grained_mod_op}, the full modular correlation function \eqref{eq:full_mod_correlation_function} then decomposes as
\begin{equation}\label{eq:full_mod_correlation_fct_coarse_grained_resolution}
    G_{\textrm{\tiny mod}} (\mathfrak{a},\mathfrak{b};t) = \sum_i p(E_i) G_{\textrm{\tiny mod}} (a(E_i),b(E_i);t,E_i)\,.
\end{equation}
Again, note the similarity to \eqref{eq:modular_correlation_function_decomposition}. 
In summary, we defined modular correlation functions in the algebra $\mathfrak{A}(L)$ as in \eqref{eq:full_mod_correlation_function} as well as a fine-grained and coarse-grained resolution as in \eqref{eq:fine_grained_modular_correlation_function_def} and \eqref{eq:coarse_grained_modular_correlation_function} respectively. 
We now investigate under which conditions these different kinds of modular correlation functions satisfy the KMS condition \eqref{eq:KMS_AQFT}. First, let us \textbf{assume} that the unresolved modular correlation function $G_{\textrm{\tiny mod}} (\mathfrak{a},\mathfrak{b};t)$ satisfies the KMS condition
\begin{equation}\label{eq:unresolved_KMS}
    G_{\textrm{\tiny mod}} (\mathfrak{a},\mathfrak{b};t+i)= \bra{\mathfrak{P}} \Delta_\mathfrak{P}^{it}\mathfrak{a}\Delta_\mathfrak{P}^{-it} \mathfrak{b}\ket{\mathfrak{P}}\,,
\end{equation}
and investigate whether the coarse- and fine-grained modular correlation functions satisfy the KMS condition as well.
Using \eqref{eq:coarse_grained_cyclic_sep_vector} and \eqref{eq:coarse_grained_mod_op}, we write
\begin{equation}
\begin{split}
    G_{\textrm{\tiny mod}} (a(E),b(E);t+i,E)&=\bra{\Psi_E}b(E)\Delta_{\Psi_E}^{i(t+i)}a(E)\Delta_{\Psi_E}^{-i(t+i)}\ket{\Psi_E}\\
    &= \int\mathrm{d}\mu(\bar{q}) \, p(\bar{q}) \mathcal{I}(E) \bra{\Psi_{\boldsymbol{\lambda}}} b(\bar{q}) \Delta_{\Psi_{\boldsymbol{\lambda}}}^{i(t+i)}a(\bar{q})\Delta_{\Psi_{\boldsymbol{\lambda}}}^{i(t+i)}\ket{\Psi_{\boldsymbol{\lambda}}}\\
    &= \bra{\mathfrak{P}} \Pi(E) \mathfrak{b}\Delta_\mathfrak{P}^{i(t+i)}\mathfrak{a}\Delta_\mathfrak{P}^{-i(t+i)}\ket{\mathfrak{P}}\\
    &= G_{\textrm{\tiny mod}} (\mathfrak{a},  \Pi(E)\mathfrak{b};t+i)\,.
\end{split}
\end{equation}
Here, we have used idempotence and the orthogonality of $\Pi(E)$ and the fact that it commutes with all operators in $\mathfrak{A}(L)$ as well as $\Delta_\mathfrak{P}$. If \eqref{eq:unresolved_KMS} is true, it is obvious that
\begin{equation}\label{eq:coarse_grained_KMS_cond}
\begin{split}
    G_{\textrm{\tiny mod}} (a(E),b(E);t+i,E)&=\bra{\mathfrak{P}} \Delta_\mathfrak{P}^{it}\mathfrak{a}\Delta_\mathfrak{P}^{-it} \Pi(E)\mathfrak{b}\ket{\mathfrak{P}}\\
    &= \int \mathrm{d}\mu(\bar{q}) \,p(\bar{q})\mathcal{I}(E) \bra{\Psi_{\boldsymbol{\lambda}}}\Delta_{\Psi_{\boldsymbol{\lambda}}}^{it}a(\bar{q})\Delta_{\Psi_{\boldsymbol{\lambda}}}^{it} b(\bar{q}) \ket{\Psi_{\boldsymbol{\lambda}}}\\
    &= \bra{\Psi_E}\Delta_{\Psi_E}^{it}a(E)\Delta_{\Psi_E}^{-it}b(E) \ket{\Psi_E}\,,
\end{split}
\end{equation}
where we again used idempotence of $\Pi(E)$. This means that given the KMS condition of $G_{\textrm{\tiny mod}} (\mathfrak{a}, \mathfrak{b};t)$, the coarse-grained modular correlation functions also satisfy their own KMS condition.
Now we turn to the fine-grained modular correlation functions. To prove the KMS condition in this case is a bit more technical, since the projectors $\Tilde{\Pi}(\bar{q})$ are not operators on $\mathfrak{H}$. We use the projectors to rewrite
\begin{equation}
\begin{split}
    G_{\textrm{\tiny mod}} (a(\bar{q}),b(\bar{q});t+i,\bar{q}) &=\frac{1}{p(\bar{q})}\bra{\mathfrak{P}}\tilde{\Pi}^\dagger(\bar{q}) b(\bar{q})\Delta_{\Psi_{\boldsymbol{\lambda}}}^{i(t+i)}a(\bar{q})\Delta_{\Psi_{\boldsymbol{\lambda}}}^{-i(t+i)}\tilde{\Pi}(\bar{q})\ket{\mathfrak{P}}\\
    &= \frac{1}{p(\bar{q})}\bra{\mathfrak{P}}\tilde{\Pi}^\dagger(\bar{q})\tilde{\Pi}(\bar{q}) \mathfrak{b}\Delta_{\mathfrak{P}}^{i(t+i)}\mathfrak{a}\Delta_{\mathfrak{P}}^{-i(t+i)}\ket{\mathfrak{P}}\,,
\end{split}
\end{equation}
where we defined $\mathfrak{a}=\int^\oplus\mathrm{d}\mu(\bar{q})\,a(\bar{q})$ and similarly $\mathfrak{b}$ and used the definitions \eqref{eq:projectors_fine_grained} and \eqref{eq:projectors_fine_grained_adjoint}. Next, note that while $\tilde{\Pi}(\bar{q})$ is not in $\mathfrak{A}(L)$, we can define the operator $\tilde{\Pi}^\dagger(\bar{q})\tilde{\Pi}(\bar{q})$ as an unbounded operator on $\mathfrak{H}$ with matrix elements given by
\begin{equation}
    \braket{\mathfrak{P}}{\tilde{\Pi}^\dagger(\bar{q})\tilde{\Pi}(\bar{q})\mathfrak{Q}}=\braket{p(\bar{q})}{q(\bar{q})}\,,
\end{equation}
which can be made arbitrarily large.
Since the fine-grained modular correlation function is finite for the operators $a(\bar{q})$ and $b(\bar{q})$ almost everywhere in $\bar{q}$, the operator $\tilde{\Pi}^\dagger(\bar{q})\tilde{\Pi}(\bar{q})$ is well defined inside the modular correlation function and we again use \eqref{eq:unresolved_KMS} to find
\begin{equation}\label{eq:fine_grained_modular_corr_KMS_condition}
\begin{split}
    G_{\textrm{\tiny mod}} (a(\bar{q}),b(\bar{q});t+i,\bar{q})&= \frac{1}{p(\bar{q})}\bra{\mathfrak{P}}\Delta_{\mathfrak{P}}^{it}\mathfrak{a}\Delta_{\mathfrak{P}}^{-it}\tilde{\Pi}^\dagger(\bar{q})\tilde{\Pi}(\bar{q}) \mathfrak{b}\ket{\mathfrak{P}}\\
    &= \bra{\Psi_{\boldsymbol{\lambda}}}\Delta_{\Psi_{\boldsymbol{\lambda}}}^{it}a(\bar{q})\Delta_{\Psi_{\boldsymbol{\lambda}}}^{-it}b(\bar{q})\ket{\Psi_{\boldsymbol{\lambda}}}\,.
\end{split}
\end{equation}
Thus, the fine-grained modular correlation function also satisfies the KMS condition.

Now, let us investigate the converse statements. Thus, we assume that the coarse-grained modular correlation functions satisfy the KMS condition as in \eqref{eq:coarse_grained_KMS_cond} and examine whether the full modular correlation function satisfies it as well. To this end, we consider a collection of disjoint $E_i$ whose union is $X$, which defines an algebra of the form \eqref{eq:coarse-grained resolution}. Using \eqref{eq:full_mod_correlation_fct_coarse_grained_resolution}, we find
\begin{equation}
\begin{split}
    G_{\textrm{\tiny mod}} (\mathfrak{a},\mathfrak{b};t+i) &= \sum_i p(E_i) G_{\textrm{\tiny mod}} (a(E_i),b(E_i);t+i,E_i)\\
    &= \sum_i p(E_i)  \bra{\Psi_E}\Delta_{\Psi_E}^{it}a(E)\Delta_{\Psi_E}^{-it}b(E) \ket{\Psi_E}\\
    &= \bra{\mathfrak{P}} \Delta_\mathfrak{P}^{it}\mathfrak{a}\Delta_\mathfrak{P}^{-it} \mathfrak{b}\ket{\mathfrak{P}}\,.
\end{split}
\end{equation}
Next, let us assume that the fine-grained modular correlation function satisfies the KMS condition as in \eqref{eq:fine_grained_modular_corr_KMS_condition} almost everywhere. Using \eqref{eq:fine_grained_modular_correlation_function}, we find again,
\begin{equation}
\begin{split}
    G_{\textrm{\tiny mod}} (\mathfrak{a},\mathfrak{b};t+i) &= \int\mathrm{d}\mu(\bar{q})\,p(\bar{q})G_{\textrm{\tiny mod}} (a(\bar{q}),b(\bar{q});t+i,\bar{q})\\
    &= \int\mathrm{d}\mu(\bar{q})\,p(\bar{q})\bra{\Psi_{\boldsymbol{\lambda}}}\Delta_{\Psi_{\boldsymbol{\lambda}}}^{it}a(\bar{q})\Delta_{\Psi_{\boldsymbol{\lambda}}}^{-it}b(\bar{q})\ket{\Psi_{\boldsymbol{\lambda}}}\\
    &= \bra{\mathfrak{P}} \Delta_\mathfrak{P}^{it}\mathfrak{a}\Delta_\mathfrak{P}^{-it} \mathfrak{b}\ket{\mathfrak{P}}\,.
\end{split}
\end{equation}
Thus, we find that if the coarse and fine-grained modular correlation functions satisfy the KMS condition, also the full modular correlation function does.

We now comment on this remarkable finding of the present manuscript. First, the validity of the KMS conditions for the symmetry-resolved modular correlation functions confirm that theory developed in this section is a full-fledged symmetry-resolved modular theory for hyperfinite algebras. Indeed, the KMS condition for the modular correlation function is a direct consequence of the Tomita-Takesaki theorem \eqref{eq:TTtheorem}. Moreover, our analyses extend and provide formal justifications to some of the results in \cite{Di_Giulio_2023}. On the one hand, the symmetry-resolved modular theory developed for type I algebras is generalized to the hyperfinite case. On the other hand, in \cite{Di_Giulio_2023}, the symmetry-resolved modular correlation functions have been computed for a two-dimensional free massless Dirac theory and, after removing the UV cutoff, the KMS condition has been observed in all the charge sectors. Although the theory developed in \cite{Di_Giulio_2023} applies to type I algebras, we expect the CFT result therein is due to the KMS condition for symmetry-resolved modular correlation functions for hyperfinite algebras. Thus, the findings of the present work formally justify the conclusions drawn in \cite{Di_Giulio_2023}.

\section{Conclusions and outlook}
\label{sec:conclusions}

In this work, we study the resolution of modular operators and modular flows into subregion-charge sectors. The main advancement consists of extending the analysis of \cite{Di_Giulio_2023}, valid for type I algebras, to hyperfinite von Neumann algebras, i.e.~type II and type III algebras.

To carry out the analysis on hyperfinite algebras, we have exploited the setup developed in \cite{Powers1967,Araki1968}, where every type II and type III algebra (up to isomorphisms) can be obtained from an appropriate limit of infinite tensor products of finite-dimensional algebras. Since the desired decomposition is based on the sectors induced by a subregion charge on an eigenstate of the total charge, in \secref{subsec:chargenorescaling}, we have introduced a charge operator in the setting of \secref{sec:AW_setup}.
Although it is a naturally defined quantity, it exhibits subtleties in the large $N$ limit, i.e.~the limit when the hyperfinite algebras are accessed. Indeed, the subregion charge does not belong to the local algebra in this regime. Inspired by ideas in the context of symmetry-resolved entanglement, in \secref{subsec:rescaling}, we considered a charge rescaled by the parameter $N$. In the large $N$ limit, this leads to a subregion charge operator, which belongs to the local algebra but is proportional to the identity.

A non-trivial symmetry resolution of the local algebra and its modular operator is obtained by using the construction developed in \secref{sec:hyeprfiniteresolved}.
The local algebra is built up by combining hyperfinite algebras obtained by the ITPFI construction. Indeed, in \secref{subsec:rescaling}, we obtained a family of hyperfinite factors labelled by the eigenvalues of the subregion charge in that algebra. Due to the rescaling of \secref{subsec:rescaling}, the subregion charge values allowed in different hyperfinite factors take continuous values, requiring the introduction of direct integrals to replace the direct sum. The construction of the algebras resolved into subregion charge sectors through direct integrals is carried out in \secref{sec:hyeprfiniteresolved}. This is the first main achievement of this manuscript.

Finally, in \secref{sec:modflow_Resolu}, we show that the modular operator and the modular flow in the algebras built in \secref{sec:hyeprfiniteresolved} can be decomposed into the charge sectors, and the analysis of \cite{Di_Giulio_2023} generalizes to the hyperfinite case. Interestingly, a symmetry resolution can also be performed for the modular correlation functions \eqref{eq:modcorrfunc_AQFT}. We find that if the modular correlation functions in the full algebra satisfy the KMS condition, then the same holds also for all the modular correlation functions in the fixed-subregion charge algebras. The decompositions obtained in \secref{sec:modflow_Resolu} and the KMS properties for the symmetry-resolved modular correlation functions are the second and third central results of this work.

As anticipated, the analyses reported in this work have connections with QFTs and holography. Indeed, to frame QFTs in a mathematically rigorous way, it is necessary to associate local type III$_1$ algebras with any causally complete region of the spacetime \cite{Haagbook}. Thus, symmetry resolution of quantities well-defined in QFT, such as relative entropies \cite{Capizzi:2021zga} and mutual information \cite{Parez:2021pgq,Ares:2022gjb}, can be framed within the approach developed in this manuscript. Moreover, as recently discussed \cite{Leutheusser:2021frk,Leutheusser:2021qhd}, type III$_1$ algebras arise in a holographic context when considering a large $N$ semiclassical theory in the bulk. When taking into account gravitational effects, the theory has to be correspondingly described in terms of type II algebras \cite{Witten:2021unn,Chandrasekaran:2022cip}.
Thus, hyperfinite algebras play a crucial role in this context. 

We expect our analysis to be of interest for using the boundary charge sectors as additional data for a refined version of the bulk reconstruction program within the AdS/CFT correspondence in the presence of charged black holes.
In this context, modular theory has proven to be insightful in the study of entanglement \cite{Ryu:2006bv} and the emergence of spacetime geometry \cite{VanRaamsdonk:2010pw}.
As discussed in \cite{Jafferis:2014lza,Jafferis:2015del}, the modular flow in the boundary theory subregions has a precise holographic dual. This leads to unravelling the connections between entanglement wedges in the bulk and the corresponding CFT regions.
In addition, the bulk locality in AdS/CFT and the bulk reconstruction program have been developed by using the toolkit of modular theory \cite{Hamilton:2006az,Kabat:2011rz,Faulkner:2017vdd,Faulkner:2018faa,DeBoer:2019kdj,Foit:2019nsr,Johnson:2022cbe}, and modular correlation functions of operators in the boundary CFT have been considered to probe the presence of quantum extremal surfaces in the bulk
\cite{Chandrasekaran:2021tkb,Chandrasekaran:2022qmq}. 
 For these reasons, our findings are potentially applicable to the AdS/CFT correspondence. As a possible strategy to incorporate symmetry resolution into this program, the analyses in \cite{Zhao:2020qmn,Weisenberger:2021eby,Zhao:2022wnp} on symmetry-resolved holographic entanglement entropy could be used.
 

Algebraic tools have been used in \cite{Magan:2021myk} to provide a formal understanding of the equipartition of entanglement in QFT. To represent the symmetry operators on local algebras, twist operators are introduced, while intertwiners are associated with reducible and irreducible representations of symmetry groups \cite{Doplicher:1972kr}. This is a promising framework that can also be applied to the symmetry resolution of the modular theory, with the ultimate goal of enriching the formal understanding gained from the results of this manuscript.
 
Finally, it is worth mentioning the potential connections between our results and the study of entanglement in gauge theories. Investigations along this line are more intricate than the study of the entanglement in the presence of global symmetries. Indeed, the presence of local gauge constraints makes the definition of spatial subregions and the consequent notion of entanglement measures very subtle. To bypass this problem, approaches based on algebraic formulation of quantum theories have been developed and applied \cite{Casini:2013rba,Casini:2014aia,Casini:2015dsg,Huerta:2018xvl,Casini:2019nmu} (see also \cite{Donnelly:2011hn,Donnelly:2016auv,Agon:2013iva,Ghosh:2015iwa,Soni:2015yga,Soni:2016ogt} for related studies).
Due to the difficulties above, understanding how to define subregion charge sectors in gauge theories is an open question. Our algebraic approach to symmetry resolution, based on the construction of fixed subregion charge operator algebras, can be successfully applied in this context and lead to progress in this field, where, despite its relevance, many directions are still to be explored.

\acknowledgments
We are grateful to Elliot Gesteau, Jonathan Karl, Thomas K{\"o}gel, Ren\'e Meyer, Ko Sanders and Leo Shaposhnik for fruitful discussions and comments. GDG is supported by the ERC Consolidator grant (number: 101125449/acronym: QComplexity). Views and opinions expressed are however those of the authors only and do not necessarily reflect those of the European Union or the European Research Council. Neither the European Union nor the granting authority can be held responsible for them. J. E.~and H. S.~are supported by the Deutsche Forschungsgemeinschaft (DFG, German Research Foundation) through
the German-Israeli Project Cooperation (DIP) grant ‘Holography and the Swampland’, as
well as under Germany’s Excellence Strategy through the Würzburg-Dresden Cluster of
Excellence on Complexity and Topology in Quantum Matter - ct.qmat (EXC 2147, project-id 390858490).

\newpage

\appendix

\section{Resolution for non-Abelian groups}\label{app:non_abelian_and_discrete_resolution}

In this appendix, we discuss generalizations to our framework, namely the resolution in the case of non-Abelian symmetry groups.
The first study on symmetry-resolved entanglement in the presence of non-Abelian symmetry groups was carried out in \cite{Calabrese:2021wvi}, while
the algebraic perspective of symmetry resolution for non-Abelian symmetry groups in the type I case was developed in \cite{Bianchi:2024aim}. Here, we extend the latter analysis for continuous spectra.

Consider a (semi-simple) Lie group describing a symmetry of the system. The Hilbert space then carries a representation of the symmetry group. This is realized by a subalgebra $\mathcal{F}_{\mathrm{sym}}\subset \mathcal{F}$, spanned by the generators $T^a$ of the Lie algebra represented on the Hilbert space. Denoting the rank of the group by $r$, we can construct $r$ independent Casimir operators $Q^a$, which commute with all operators in $\mathcal{F}_{\mathrm{sym}}$.
Similar to the discussion in \secref{sec:modular_theory}, the Hilbert space will decompose into superselection sectors. In the non-Abelian case, these are labelled by the set of eigenvalues $\boldsymbol{q}$ of the Casimir operators constructed by contracting symmetric invariant tensors,
\begin{equation}
    \mathcal{H}= \bigoplus_{\boldsymbol{q}}\mathcal{H}_{\boldsymbol{q}}\,.
\end{equation}
We define the algebra of invariant operators as
\begin{equation}
    \pi_{\boldsymbol{q}}(\mathcal{A})\equiv\pi_{\boldsymbol{q}}( (\mathcal{F}_{\mathrm{sym}})')= P_{\boldsymbol{q}}\{a\in \mathcal{F} \,\vert\, [a,T]=0\,\forall T\in \mathcal{F}_{\mathrm{sym}} \} P_{\boldsymbol{q}}\,,
\end{equation}
where $ \pi_{\boldsymbol{q}}$ denotes the projector into the superselection sector labelled by $\boldsymbol{q}$. As in the discussion in the main text, we will in the following drop the projector and implicitly assume that the algebra is represented on the vacuum sector $\boldsymbol{q}=0$. 
The invariant algebra has a centre that is generated by the Casimir operators $Q^a$. We emphasize that, by definition, the centre is an Abelian algebra even if $\mathcal{F}_{\mathrm{sym}}$ is non-Abelian.

Similarly, we can define the invariant algebra associated with a subregion $V$ by
\begin{equation}
    \mathcal{A}(V) = \{a\in \mathcal{F}(V) \,\vert\, [a,T]=0\,\forall T\in \mathcal{F}_{\mathrm{sym}} \}\,.
\end{equation}
We assume that the algebra still has a centre generated by operators $Q^a_V\in \mathcal{A}(V)$. This is the case for example if we consider operators that are local in the sense that they can be obtained by an integral over a density. Formally, this is realized by representing, in the group sense, the Casimir operators not only in $\mathcal{F}$ but also in $\mathcal{F}(V)$ (and consequently on $\mathcal{A}(V)$).
We collectively denote the eigenvalues of the restricted Casimir operators $Q^a_V$ by $\bar{\boldsymbol{q}}$. By successively applying the spectral theorem, we can decompose the invariant algebra into successive direct integrals
\begin{equation}
    \mathcal{A}(V) = \int_{\sigma(T_1)}^\oplus \mathrm{d}\mu(q_1)\dots \int_{\sigma(T_2)}^\oplus \mathrm{d}\mu(q_2) \,\mathcal{A}_{q_1,\dots,q_r}\,.
\end{equation}
Similarly, the Hilbert space may be decomposed as
\begin{equation}
    \mathcal{H} = \int_{\sigma(T_1)}^\oplus \mathrm{d}\mu(q_1)\dots \int_{\sigma(T_2)}^\oplus \mathrm{d}\mu(q_2) \,\mathcal{H}_{q_1,\dots,q_r}\,.
\end{equation}
The algebra $\mathcal{A}_{q_1,\dots,q_r}$ is still a von Neumann algebra acting on the Hilbert space $\mathcal{H}_{q_1,\dots,q_r}$.


\section{Convergence of total and local magnetization operators}\label{app:convergence_ZN_ZNL_ZNR}

In this appendix, we provide detailed calculations on the large $N$ limit of the operators defined in \eqref{eq:projectors_finite_N}, \eqref{eq:full_mag_finite_N}, and \eqref{eq:subregion_magnetization_sequence_defintion}.

\paragraph{Convergence of $Z^{(N)}$:} First, consider an arbitrary state in the pre-Hilbert space $\ket{\theta}=\theta_1\otimes\dots\otimes \theta_k\otimes K_{2,\lambda_{k+1}}\otimes\dots\in \tilde{\mathcal{H}}$. For any $N>k$, we have
\begin{equation}
    (z_L^N+z_R^N)\ket{\theta}=0\,,
\end{equation}
as discussed below \eqref{eq:eigenstate_AW}. Thus, we have
\begin{equation}
    Z^{(N)}\ket{\theta}=Z^{(k)}\ket{\theta}\,,
\end{equation}
and for $N,M>k$,
\begin{equation}
    (Z^{(N)}-Z^{(M)})\ket{\theta}=0\,.
\end{equation}
Thus, \eqref{eq:cauchy_seq_condition} is satisfied for $\tilde{\mathcal{H}}$. Since $\tilde{\mathcal{H}}$ is dense in $\mathcal{H}$, $Z^{(N)}$ also converges on $\mathcal{H}$. We can denote the limit $Z=\lim_{N\to\infty} Z^{(N)}$.

\paragraph{Non-convergence of $Z_L^{(N)},Z_R^{(N)}$:} Although the sum of $Z_L^{(N)}$ and $Z_R^{(N)}$ converges, they individually do not. To see this, we observe that the sequence fails to converge on the cyclic separating vector \eqref{eq:cyclicseparating AW}. Indeed,
\begin{equation}\label{eq:sub_mag_non_convergence}
    ||(Z_L^{(N)}-Z_L^{(N-1)})\ket{\Psi_{\boldsymbol{\lambda}}}||^2=||z_L^N \ket{\Psi_{\boldsymbol{\lambda}}}||^2=\frac{1}{4}\mathrm{tr}\left(K_{2,\lambda_N}^\dagger \sigma_z^\dagger \sigma_z K_{2,\lambda_N}\right)=\frac{1}{4}\,.
\end{equation}
Here, in the first step, we used definition \eqref{eq:subregion_magnetization_sequence_defintion}. This equation violates the necessary condition for convergence, since this expression is constant and does not tend to $0$ as $N\to \infty$. In particular, \eqref{eq:strong_convence_criterion} cannot be satisfied for the cyclic separating vector. The same argument also holds for $Z_R^{(N)}$.
We conclude that the sequences $Z_L^{(N)},Z_R^{(N)}$ do not converge in the full Hilbert space $\mathcal{H}$.
At this point, it is natural to ask if there are bounded functions of $Z_L^{(N)},Z_R^{(N)}$ that converge. This is interesting since the projectors \eqref{eq:projectors_finite_N} are examples of these bounded functions.

\paragraph{Non-convergence of bounded functions of $Z_L^{(N)},Z_R^{(N)}$:}

Consider the basis of bounded functions of $Z_L^{(N)},Z_R^{(N)}$ spanned by exponentials of the form
\begin{equation}
    b_\alpha^{(N)}\equiv e^{i\alpha Z_L^{(N)}}\in\mathcal{F}(L)\,.
\end{equation}
Note that for finite $N$ operators of this form truncate in their tensor product expansion such that after the first $N$ factors, the tensor product consists of products of the $2\times 2$ unit matrix, i.e.~$b_\alpha^{(N)}\in \tilde{\mathcal{F}}(L)$. Since the spectrum \eqref{eq:mag_finite_N_spectrum} is half-integer spaced for finite $N$, it suffices to consider $\alpha\in (-4\pi,4\pi)$. Note that $b_0^{(N)}=\mathrm{id}_L$. Moreover, we may check that the exponentials are unitary, i.e.~that for all $\ket{\theta}\in\mathcal{H}$
\begin{equation}\label{eq:bounded_fct_unitary}
    ||b_{\alpha}^{(N)} \ket{\theta}||^2=||\ket{\theta}||^2\,.
\end{equation}
Moreover, we may write
\begin{equation}
    b_{\alpha}^{(N+1)}=e^{i\alpha Z_L^{(N+1)}}=e^{i\alpha Z_L^{(N)}+i\alpha z_L^{N+1}}=e^{i\alpha Z_L^{(N)}} e^{i\alpha z_L^{N+1}}\,.
\end{equation}
Here, we used the fact that the operators $z_L^i$ commute. In the following discussion, it is advantageous to rewrite the right exponential in the above equation,
\begin{equation}\label{eq:pauli_formula}
    e^{i\alpha z_L^{N+1}}=\cos\frac{\alpha}{2}\mathrm{id}_L+2i\sin\frac{\alpha}{2} z_L^{N+1}\,,
\end{equation}
which can be verified using the multiplication rules of the Pauli matrices. We then find
\begin{equation}
\begin{split}
    ||(b_\alpha^{(N+1)}-b_\alpha^{(N)})\ket{\Psi_{\boldsymbol{\lambda}}}||^2&=||(e^{i\alpha Z_L^{(N+1)}}-e^{i\alpha Z_L^{(N)}})\ket{\Psi_{\boldsymbol{\lambda}}}||^2\\
    &= ||(\cos\frac{\alpha}{2}-1+2i \sin\frac{\alpha}{2}z_L^{N+1})\ket{\Psi_{\boldsymbol{\lambda}}}||^2\,,
\end{split}
\end{equation}
where we used \eqref{eq:bounded_fct_unitary} and \eqref{eq:pauli_formula}. Using the fact that $z_L^{N+1}$ is hermitian as well as \linebreak\hbox{$||z_L^N \ket{\Psi_{\boldsymbol{\lambda}}}||^2=\frac{1}{4}$}, we find
\begin{equation}
    ||(\cos\frac{\alpha}{2}-1+2i \sin\frac{\alpha}{2}z_L^{N+1})\ket{\Psi_{\boldsymbol{\lambda}}}||^2=2-2\cos\frac{\alpha}{2}\,.
\end{equation}
As in \eqref{eq:sub_mag_non_convergence}, we find that the difference is independent of $N$. The only way the $b_\alpha^{(N)}$ to converge is thus
\begin{equation}
    2-2\cos\frac{\alpha}{2}\overset{!}{=} 0\,,\quad \Leftrightarrow\quad \alpha=4\pi \mathbb{Z}\,.
\end{equation}
The only solution in the domain $(-4\pi,4\pi)$ is $\alpha=0$, which corresponds to the operator $\mathrm{id}_L$. The same argument also holds for bounded functions of $Z_R^{(N)}$. Therefore, even bounded functions of the subregion magnetization fail to converge on the full Hilbert space. In particular, this includes the projectors \eqref{eq:projectors_finite_N}. 
Since $Z_L^{(N)}$, $Z_R^{(N)}$ and any of their bounded functions do not belong to the hyperfinite algebras $\mathcal{F}(L)$ and $\mathcal{F}(R)$ respectively, we cannot use these operators to define an invariant local hyperfinite algebra. Thus, the starting point for the symmetry resolution we aim for cannot be realized in this way.

We conclude this section with a comment on this issue. In the case where the algebras $\mathcal{F}(L)$ and $\mathcal{F}(R)$ become type III$_\lambda$ factors, there is also an interesting connection of the results above to modular theory. As it is possible to prove using ITPFIs, in the type III$_\lambda$ case, i.e.~$\lambda_i=\lambda$, the operator $Z^{(N)}$ is proportional to the modular Hamiltonian for the algebra of operators that only have identity matrix factors after $N$ tensor products, i.e.
\begin{equation}
    \{a_1\otimes\dots \otimes a_N\otimes \1\otimes \dots\,\}.
\end{equation}
In the limit $N\to\infty$, $Z^{(N)}$ converges to the modular Hamiltonian and $\lim_{N\to\infty} Z^{(N)}\in\mathcal{F}$, as expected. The operators $Z_L^{(N)}$ and $Z_R^{(N)}$ then correspond to the respective one-sided modular Hamiltonians. The non-existence of their limit is thus consistent with the fact that the one-sided modular Hamiltonian in type III algebras is not an element of the respective subregion algebra.
In this case, the algebra of invariant operators localized in $L$, i.e.~operators that commute with the modular Hamiltonian, is called the centralizer algebra. As shown in \cite{Takesaki1979}, the centralizer of a type III$_\lambda$ algebra is a {\it factor} of type II$_1$. Importantly, this means that the algebra does not have a non-trivial centre, which hampers the desired decomposition of the algebra.

\section{Convergence properties of rescaled charge in type \texorpdfstring{III$_1$}{III1} algebras}\label{app:convergence_rescaled_carge_III_1}

In this appendix, we discuss the convergence properties of \eqref{eq:III_1} as $N$ becomes large, which is highly dependent on the sequence $\lambda_i$. We will illustrate this fact by constructing exemplary sequences for which \eqref{eq:III_1} converges and for which it does not converge.

\begin{figure}
    \centering
    \includegraphics[width=0.5\linewidth]{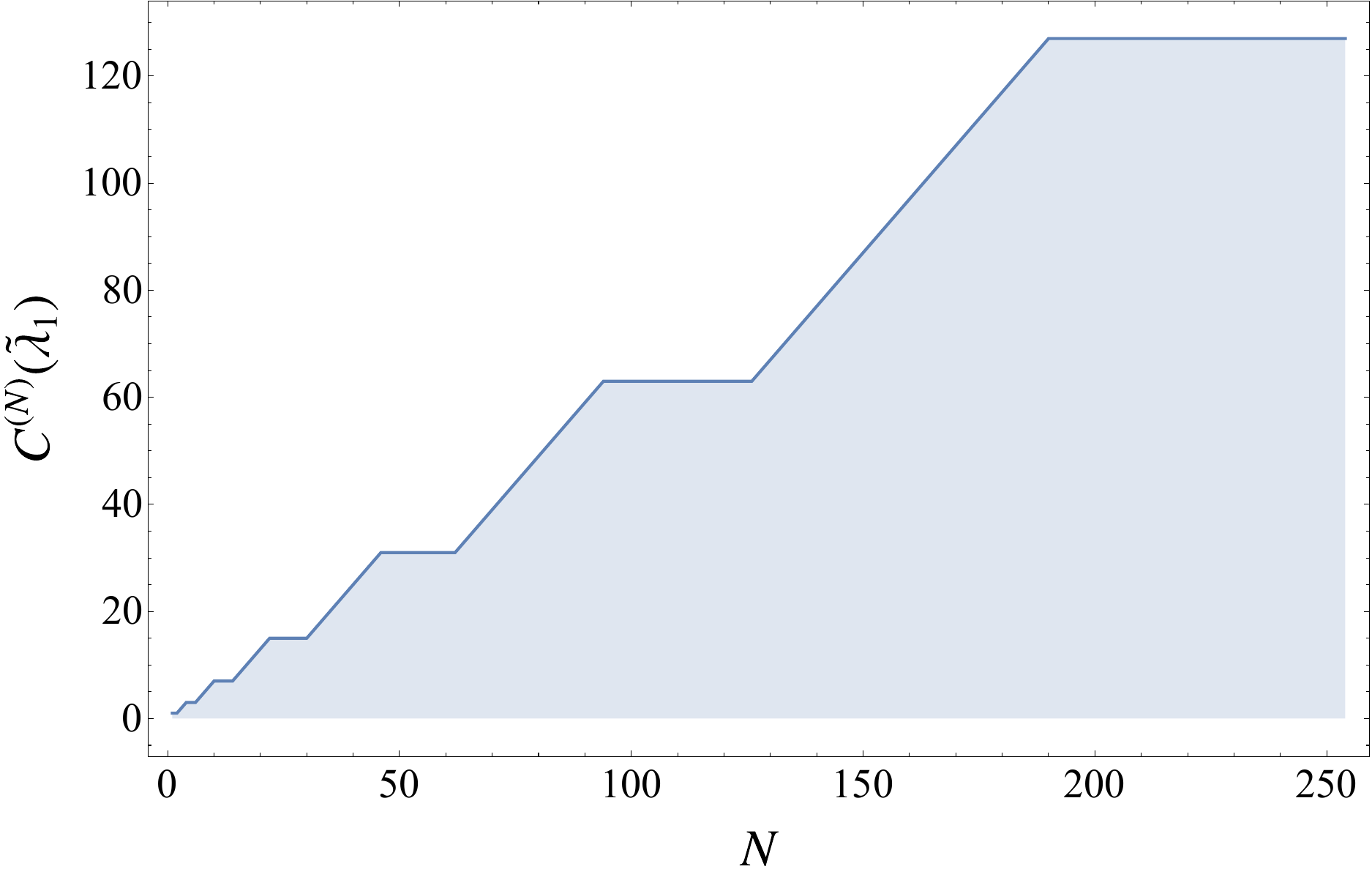}
    \caption{$C^{(N)}({\tilde{\lambda}_1})$ associated with the sequence \eqref{eq:seq_divergent} plotted as a function of $N$. }
    \label{fig:stepped_count}
\end{figure}

\paragraph{Convergence:} Consider the sequence
\begin{equation}\label{eq:example_convergent_rescaled_mag}
    \lambda_i = \begin{cases}
        \tilde{\lambda}_1, & i\text{ even}\\
        \tilde{\lambda}_2, & i\text{ odd}
    \end{cases}\,,
\end{equation}
i.e.~the sequence oscillates between two values. Clearly, this sequence is not convergent as $i$ approaches infinity, because it has more than one accumulation point. Thus, \eqref{eq:example_convergent_rescaled_mag} defines an algebra of type III$_1$. The right side of \eqref{eq:III_1} in this case reads
\begin{equation}
\begin{split}
    \frac{1}{2}\frac{1}{N}\sum_{i=1}^N \frac{1-\lambda_i}{1+\lambda_i} &= \frac{1}{2}\frac{1}{N}\sum_{i\text{ even}}\frac{1-\tilde{\lambda}_1}{1+\tilde{\lambda_1}}+\frac{1}{2}\frac{1}{N}\sum_{i\text{ odd}}\frac{1-\tilde{\lambda}_2}{1+\tilde{\lambda_2}}\\
    &=\frac{1}{2}
    \begin{cases}
        \frac{1}{2} \frac{1-\tilde{\lambda}_1}{1+\tilde{\lambda_1}} +\frac{1}{2} \frac{1-\tilde{\lambda}_2}{1+\tilde{\lambda_2}}, & N \text{ even}\\
        \frac{1}{2} \frac{1-\tilde{\lambda}_1}{1+\tilde{\lambda_1}} +\frac{1}{2} \frac{1-\tilde{\lambda}_2}{1+\tilde{\lambda_2}}+\frac{1}{N}\frac{1-\tilde{\lambda}_1}{1+\tilde{\lambda_1}} , & N \text{ odd}
    \end{cases}\,.
\end{split}
\end{equation}
Thus, we observe that
\begin{equation}
    \lim_{N\to\infty} \frac{1}{2}\frac{1}{N}\sum_{i=1}^N \frac{1-\lambda_i}{1+\lambda_i} = \frac{1}{4} \frac{1-\tilde{\lambda}_1}{1+\tilde{\lambda_1}} +\frac{1}{4} \frac{1-\tilde{\lambda}_2}{1+\tilde{\lambda_2}}\,.
\end{equation}
This can be straightforwardly, though laboriously, generalized in several ways. For example, instead of the simple behaviour in \eqref{eq:example_convergent_rescaled_mag}, the same convergence behaviour of \eqref{eq:III_1} can be generated if the sequence $\lambda_i$ is made up of two convergent subsequences for even and odd $i$, converging to $\tilde{\lambda}_1$ and $\tilde{\lambda}_2$ respectively. Moreover, an analogous behaviour can be observed for multiple convergent subsequences. Let $\iota_k$ be the index set of one of the convergent subsequences with limit $\tilde{\lambda}_k$. We denote the number of members in each convergent subsequence by
\begin{equation}
    C^{(N)}({\tilde{\lambda}_k})=|\{i\in \iota_p|i\leq N\}|\,,
\end{equation}
where $|\cdot|$ denotes the cardinality of a finite set. If we assume that 
\begin{equation}\label{eq:relative_number}
    \lim_{N\to\infty} \frac{C^{(N)}({\tilde{\lambda}_k})}{N} =\mu_k\,,
\end{equation}
exists for all $\tilde{\lambda}_k$, we can write in general
\begin{equation}
    \frac{1}{2}\frac{1}{N}\sum_{i=1}^N \frac{1-\lambda_i}{1+\lambda_i} = \frac{1}{2} \sum_{k=1}^p \mu_k \frac{1-\tilde{\lambda}_k}{1+\tilde{\lambda}_k}\,.
\end{equation}
However, the existence of the limit in \eqref{eq:relative_number} is an assumption which can be easily broken, as we will now describe with an example.
\paragraph{Non-convergence:} Consider now the sequence
\begin{equation}\label{eq:seq_divergent}
    \lambda_i = \{\tilde{\lambda}_1, \tilde{\lambda}_2, \tilde{\lambda}_1, \tilde{\lambda}_1, \tilde{\lambda}_2,\tilde{\lambda}_2, \tilde{\lambda}_1, \tilde{\lambda}_1, \tilde{\lambda}_1, \tilde{\lambda}_1, \tilde{\lambda}_2,\tilde{\lambda}_2,\tilde{\lambda}_2,\tilde{\lambda}_2,\dots\}\,,
\end{equation}
where the number in each group grows exponentially. In this case, the two subsequences are still perfectly convergent. As a function of $N$, the number $C^{(N)}({\tilde{\lambda}_k})$ is a stepped function with exponentially increasing step width. Consider for example $C^{(N)}({\tilde{\lambda}_1})$, shown in \figref{fig:stepped_count}. 
The $n$-th upper kink point occurs at
\begin{equation}
    N_{\mathrm{max}} = \sum_{j=1}^n 2^j +2^n\,,
\end{equation}
where
\begin{equation}
    C^{(N_{\mathrm{max}})}({\tilde{\lambda}_1}) = \frac{1}{2}\sum_{j=1}^n 2^j+2^n\,.
\end{equation}
The $n$-th lower kink point occurs at
\begin{equation}
    N_{\mathrm{min}} = \sum_{j=1}^n 2^j\,,
\end{equation}
where
\begin{equation}
    C^{(N_{\mathrm{min}})}({\tilde{\lambda}_1}) = \frac{1}{2}\sum_{j=1}^{n-1} 2 ^j +2^{n-1}\,.
\end{equation}
The fraction $C^{(N)}({\tilde{\lambda}_1})/N$ oscillates between $\frac{1}{2}$ and
\begin{equation}
    \frac{C^{(N_{\mathrm{max}})}({\tilde{\lambda}_1})}{ N_{\mathrm{max}} } = \frac{\frac{1}{2}\sum_{j=1}^n 2^j+2 ^n}{\sum_{j=1}^n 2^j+2^n}\,.
\end{equation}
We find that as $n\to\infty$, this ratio approaches $\frac{2}{3}$. Therefore, for the sequence \eqref{eq:seq_divergent}, the relative number of terms does not converge. By extension, we find that also the rescaled magnetization in \eqref{eq:III_1} is not convergent. We conclude that for type III$_1$ algebras, the rescaled magnetization, in general, does not exist in the large $N$ limit.

\bibliographystyle{nb.bst}
\bibliography{literature}

\end{document}

%% file: Figures/AWconstruction.pdf_tex
\begingroup%
  \makeatletter%
  \providecommand\color[2][]{%
    \errmessage{(Inkscape) Color is used for the text in Inkscape, but the package 'color.sty' is not loaded}%
    \renewcommand\color[2][]{}%
  }%
  \providecommand\transparent[1]{%
    \errmessage{(Inkscape) Transparency is used (non-zero) for the text in Inkscape, but the package 'transparent.sty' is not loaded}%
    \renewcommand\transparent[1]{}%
  }%
  \providecommand\rotatebox[2]{#2}%
  \newcommand*\fsize{\dimexpr\f@size pt\relax}%
  \newcommand*\lineheight[1]{\fontsize{\fsize}{#1\fsize}\selectfont}%
  \ifx\svgwidth\undefined%
    \setlength{\unitlength}{77.55606404bp}%
    \ifx\svgscale\undefined%
      \relax%
    \else%
      \setlength{\unitlength}{\unitlength * \real{\svgscale}}%
    \fi%
  \else%
    \setlength{\unitlength}{\svgwidth}%
  \fi%
  \global\let\svgwidth\undefined%
  \global\let\svgscale\undefined%
  \makeatother%
  \begin{picture}(1,1.50469548)%
    \lineheight{1}%
    \setlength\tabcolsep{0pt}%
    \put(0,0){\includegraphics[width=\unitlength,page=1]{AWconstruction.pdf}}%
    \put(0.50843787,0.39678452){\makebox(0,0)[lt]{\lineheight{1.25}\smash{\begin{tabular}[t]{l}$\otimes$\end{tabular}}}}%
    \put(0.50843787,0.70210559){\makebox(0,0)[lt]{\lineheight{1.25}\smash{\begin{tabular}[t]{l}$\otimes$\end{tabular}}}}%
    \put(0.50843787,0.99203903){\makebox(0,0)[lt]{\lineheight{1.25}\smash{\begin{tabular}[t]{l}$\otimes$\end{tabular}}}}%
    \put(0,0){\includegraphics[width=\unitlength,page=2]{AWconstruction.pdf}}%
    \put(0.28637119,0.05453988){\makebox(0,0)[lt]{\lineheight{1.25}\smash{\begin{tabular}[t]{l}$L$\end{tabular}}}}%
    \put(0.79806601,0.05453988){\makebox(0,0)[lt]{\lineheight{1.25}\smash{\begin{tabular}[t]{l}$R$\end{tabular}}}}%
    \put(0.0599175,0.23972814){\makebox(0,0)[lt]{\lineheight{1.25}\smash{\begin{tabular}[t]{l}$1$\end{tabular}}}}%
    \put(0.0599175,0.54065255){\makebox(0,0)[lt]{\lineheight{1.25}\smash{\begin{tabular}[t]{l}$2$\end{tabular}}}}%
    \put(0.0599175,0.84157703){\makebox(0,0)[lt]{\lineheight{1.25}\smash{\begin{tabular}[t]{l}$3$\end{tabular}}}}%
    \put(0.0599175,1.3274962){\makebox(0,0)[lt]{\lineheight{1.25}\smash{\begin{tabular}[t]{l}$N$\end{tabular}}}}%
    \put(0.50843787,1.20014777){\makebox(0,0)[lt]{\lineheight{1.25}\smash{\begin{tabular}[t]{l}$\otimes$\end{tabular}}}}%
    \put(0.07090807,1.10488587){\makebox(0,0)[lt]{\lineheight{1.25}\smash{\begin{tabular}[t]{l}$\vdots$\end{tabular}}}}%
    \put(0.52959155,1.09389528){\makebox(0,0)[lt]{\lineheight{1.25}\smash{\begin{tabular}[t]{l}$\vdots$\end{tabular}}}}%
    \put(0,0){\includegraphics[width=\unitlength,page=3]{AWconstruction.pdf}}%
  \end{picture}%
\endgroup%